\documentclass[conference]{IEEEtran}

\usepackage{xspace}
\usepackage{mathtools}   

\usepackage{graphicx}
\usepackage{tikz}

\usepackage{pgfplots}
\pgfplotsset{compat=1.15}
\pgfplotsset{
  grid style = {
    dash pattern = on 1mm off 2mm,
    line cap     = round,
    black!15,
    line width   = .2pt
  },
}
\pgfplotsset{
  base/.style={
    title style         = {at={(0.45,1.1)}},
    scale only axis,
    width               = 7in,
    height              = 3.5in,
    enlarge x limits    = 0.11,
    tick align          = inside,
    x label style       = {at = {(0.65,-0.15)}},
    y tick label style  = {/pgf/number format/assume math mode},
  },
}
\pgfplotsset{
  nnc/.style={
    ybar       = .75pt,
    bar width  = 11pt,
    nodes near coords,
  }
}

\usepackage{multirow}
\usepackage{multicol}
\usepackage{diagbox}   

\usepackage{enumitem}
\setlist{nosep}

\usepackage{xcolor}

\definecolor{nblue}{RGB}{46,134,193}
\definecolor{ngr}{RGB}{142,68,173}
\definecolor{col1}{HTML}{7D3C98}
\definecolor{col2}{HTML}{BD910F}
\definecolor{nb}{HTML}{00BBFF}

\usepackage[normalem]{ulem}
\newcommand\Sout{\textcolor{red}\bgroup
  \markoverwith{\textcolor{red}{\rule[.5ex]{4pt}{0.4pt}}}\ULon}

\usepackage[linesnumbered,ruled,vlined]{algorithm2e}

\SetCommentSty{mycommfont}

\usepackage{hyperref}

\newcommand{\Omit}[1]{}

\newcommand{\RNum}[1]{\uppercase\expandafter{\romannumeral #1\relax}}

\newcommand{\remove}[1]{}

\newcommand{\Wset}{\textit{Wset}}
\newcommand{\writeset}{\textit{write-set}}
\newcommand{\Rset}{\textit{Rset}}
\newcommand{\readset}{\textit{read-set}}
\newcommand{\Dset}{\textit{Dset}}
\newcommand{\accessset}{\textit{access-set}}

\newcommand{\Read}{\textit{read}}

\newcommand{\Write}{\textit{write}}

\newcommand{\TryC}{\textit{tryC}}

\newcommand{\txns}{\textit{txns}}

\newcommand{\BSTM}{Block-STM}
\newcommand{\preset}{preset}
\newcommand{\Preset}{Preset}

\newcommand{\presetSerialization}{preset serialization}

\newcommand{\rwOblivious}{read-write-oblivious}
\newcommand{\RWO}{Read-Write-Oblivious}
\newcommand{\RwOblivious}{Read-write-oblivious}
\newcommand{\rwAware}{read-write-aware}
\newcommand{\RwAware}{Read-write-aware}
\newcommand{\RWA}{Read-Write-Aware}

\newcommand{\dApp}{dApp}

\newcommand{\ignore}[1]{}

\newcommand*\circled[1]{%
  \tikz[baseline=(char.base)]{
    \node[shape=circle,draw,inner sep=.25pt] (char) {#1};}}

\usepackage{amsthm}
\tikzset{every picture/.style={scale=.65}, font = \scriptsize}

\usepackage{cleveref}
\usepackage{comment}
\usepackage{cite}

\theoremstyle{definition}
\newtheorem{definition}{Definition}

\usepackage[font=footnotesize,labelfont=bf,labelsep=period,skip=2pt]{caption}
\title{Blockchain Transaction Conflicts: A Historical Perspective}

\author{
\IEEEauthorblockN{Parwat Singh Anjana}
\IEEEauthorblockA{
Supra Research \\
p.anjana@supra.com
}
\and
\IEEEauthorblockN{Srivatsan Ravi}
\IEEEauthorblockA{
Supra Research and University of Southern California \\
s.ravi@supra.com 
}
\and
\IEEEauthorblockN{Maurice Herlihy}
\IEEEauthorblockA{
Brown University\\
maurice.herlihy@gmail.com
}
}

\begin{document}
\maketitle
\begin{abstract}
    This paper presents a comprehensive analysis of historical data across two popular blockchain networks: Ethereum and Solana. Our study focuses on two key aspects: transaction conflicts and the maximum theoretical parallelism within historical blocks. We aim to quantify the degree of transaction parallelism and assess how effectively it can be exploited by systematically examining block-level characteristics, both within individual blocks and across different historical periods. 
     In particular, this study is the first of its kind to leverage historical transactional workloads to evaluate conflict patterns. By offering a structured approach to analyzing these conflicts, our research provides valuable insights and an empirical basis for developing more efficient parallel execution techniques for smart contracts in the Ethereum and Solana. 
     Our empirical analysis reveals that historical Ethereum blocks frequently achieve high independence, with over 50\% independent transactions in more than 50\% of blocks, while, on average, Solana blocks contain longer conflict chains $\sim$58\%, compared to $\sim$18\% in Ethereum, reflecting fundamentally different parallel execution dynamics.
    
\end{abstract}

\begin{IEEEkeywords}
Blockchains, Parallel Execution, Ethereum, Solana, Empirical Analysis
\end{IEEEkeywords}

\section{Introduction \label{sec:intro}}
Blockchain virtual machines (VMs) execute a \emph{block} of user-defined \emph{transactions}, each performing a sequence of \Read{s} and \Write{s} on the global state. Each VM takes as input a block consisting of $n$ transactions and a \emph{\preset{}} total order $T_1\rightarrow T_2\ldots\rightarrow T_n$. The VM attempts to execute these transactions \textit{in parallel} in such a way that the blockchain's final state is identical to the state obtained by executing those transactions \emph{sequentially}. All VMs are given the same \presetSerialization{} order~\cite{blockstm,Anjana2025BTMSSS} needed to guarantee that the independent VMs agree on the block's effects.

Different VMs utilize different execution models. Ethereum and Solana represent two distinct approaches. Ethereum processes transactions sequentially, without prior knowledge of how transactions will access the blockchain state. 
Instead, state access patterns become known only at execution time,
a strategy we call the~\emph{\rwOblivious{} model}.
By contrast, Solana requires clients to pre-declare state access patterns,
a strategy we call the  \emph{\rwAware{} model}.
Naturally, VMs based on these different models require different run-time structures.


The degree to which VMs can execute transactions in parallel depends on how those
transactions interact.
A \emph{conflict} occurs when two or more transactions access the same
storage slot or account address,
and at least one performs an update operation.
Conflicts limit concurrency:
conflicting transactions must be executed in the block's \preset{} order,
but \emph{independent} (non-conflicting) transactions may be executed in any convenient order.

This study analyzes the potential and the limits of parallelization
for transaction executions in blockchain VMs.
These are empirical questions.
It is possible, in principle,
to parallelize executions for transactions with few conflicts,
but no scheme, however clever,
can parallelize executions for transactions with many conflicts.
Which is it?
The blockchains themselves are a rich source of historical data,
recording millions of transactions over the course of years.
This study analyzes the historical distribution of conflicts and state
access patterns in Solana and Ethereum as they evolve over time, under both normal and exceptional loads.

There are several ways to evaluate potential parallelism.
The \emph{longest conflict chain} (or critical path, or span)
is one of the principal measures determining the maximum possible speedup
achievable through parallel execution~\cite{BlumofeCilk,BlumofeWorkStealing,HM93,GraceBerger,Anjana2025BTMSSS}.
Related measures include the number of \emph{conflicting transactions},
total \emph{conflicts},
and the number of \emph{conflict~families},
groups of transactions mutually dependent on shared states.

Potential conflicts can be identified through static analysis or simulation.
Actual conflicts can be identified at run-time by tracking access patterns,
either pessimistically (via locks) or optimistically (via roll-back and retry).


Prior research on characterizing parallel transaction execution in blockchains has largely focused on Ethereum and follows three key directions: static analysis-based \emph{conflict prediction}, \emph{conflict-aware scheduling}, and \emph{empirical analysis of transaction conflicts}. 
Static analysis helps identify potential conflicts upfront at the code level~\cite{chahoki2025static}, while scheduling systems demonstrate how conflict information can be efficiently utilized for deterministic parallel execution~\cite{iBTM_AFT_2025,chahoki2025conthereum}; empirical analysis, in turn, reveals real-world conflict patterns that bound the maximum achievable speedup (e.g., skewed workload, hot spots, conflict chains, etc.)~\cite{saraph2019empirical,biton2025conflicts}. However, the literature remains largely Ethereum-centric and often evaluates either a specific execution strategy (speculation or a particular scheduler) or a single layer of dependence (call-level or coarse-grained), with limited cross-platform analysis on how execution models and state-access semantics shift conflict patterns and parallelism in practice. Moreover, ecosystem dependency analysis~\cite{jin2025dependency} demonstrates distribution but does not directly quantify the fine-grained read-write conflicts that determine parallelism within a block.

\noindent\textbf{Contributions:} This study provides an empirical foundation  for understanding  parallel execution {across} execution models: we analyze transaction conflicts at scale on both Ethereum and Solana, and quantify conflict derived bounds on parallelism under consistent, execution-oriented traces. Relative to static analysis~\cite{chahoki2025static}, we focus on realized on-chain behavior and its measurable concurrency constraints; relative to scheduling systems~\cite{chahoki2025conthereum}, we provide workload evidence and structural metrics that can guide (and stress-test) scheduler design rather than proposing a new scheduler; relative to speculative analysis~\cite{saraph2019empirical} and Ethereum-only conflict studies~\cite{biton2025conflicts}, we broaden the empirical scope to a second major execution paradigm and emphasize comparative conflict structure and speedup bounds; and relative to ecosystem dependency risk analysis~\cite{jin2025dependency}, we connect contract interaction complexity to transaction-level dependencies by directly measuring conflicts that enforce serialization at execution time. 

We analyze blocks from Ethereum and Solana with \rwOblivious{} and \rwAware{} execution models to understand their respective conflict distributions and 
also present a comparative study of conflict patterns to empirically assess execution models. 
Our findings offer insights into the \emph{ground truth} for the maximum parallelism
, constrained by inherent conflicts within historical blocks. 
We conjecture that the best possible parallel execution will need to leverage conflict relationships \emph{a priori} to achieve the best possible parallelism. 
Our analysis reveals that more than 50\% of the blocks consistently contain over 50\% independent transactions, with over 94\% exceeding the 40\% threshold, 
indicating a growing opportunity for parallel execution in Ethereum. In contrast, \emph{recent historical blocks} in Solana exhibit significantly longer conflict chains ($\sim$36\% of block size vs. Ethereum's $\sim$8\%), fewer conflict families ($\sim$39 vs. $\sim$130) and only $\sim$7\% independent transactions compared to Ethereum's $\sim$64\%, highlighting the dense dependency patterns in Solana. 


\noindent\textbf{Paper Outline:} \Cref{sec:background} presents the background and motivation. 
The system model and formal conflict definitions are provided in \Cref{sec:sm}. \Cref{sec:conf_analysis} introduces the metrics 
and describes the data extraction process. The empirical analysis is presented in \Cref{sec:eth_analysis} for Ethereum and \Cref{sec:sol_analysis} for Solana. A comparative study of 
Ethereum and Solana is provided in \Cref{sec:eth_vs_solana}. Finally, we conclude in \Cref{sec:conc}. 

\section{Motivation\label{sec:background}}
With the evolution of blockchains, parallel execution becomes critical for scalability and higher throughput; however, its efficiency is fundamentally limited by transaction conflicts at runtime. 
As discussed in \Cref{sec:intro}, Ethereum and Solana represent two distinct execution models. The former employs a single-threaded, sequential execution model ({\rwOblivious{}}), while the latter executes transactions in parallel ({\rwAware{}}) based on explicit read-write sets. 

In the {\rwOblivious{}} model, the execution engine dynamically detects conflicts at runtime, requiring STM (software transactional memory) or OCC (optimistic concurrency control) based mechanisms. To preserve determinism across the network, parallel execution must satisfy the predefined serialization order. Transactions are executed speculatively in parallel under the assumption of independence; however, upon conflict detection, they are rolled back and re-executed. This incurs execution overhead, particularly under high-contention workloads. As a result, in this model, parallelism and overall performance are limited by \preset{} serialization and wasted work induced by aborts. 

In the \rwAware{} model, the read-write sets are known prior to execution. This enables static construction of dependencies
, allowing for informed scheduling and conflict-aware parallel execution, resulting in an 
optimal or near-optimal execution of non-conflicting transactions across multiple threads. 
This approach results in deterministic execution with minimal rollback overhead. 
However, accurately determining read-write sets incurs additional computational overhead, particularly for contracts with complex or data-dependent behavior, while explicit access lists inflate transaction size and bandwidth.

\begin{figure}[!b]
    \vspace{-.5cm}
    \centering
    \scalebox{.8}{\input{figs/ethereum_solana_exe}}
    \vspace{-.45cm}
    \caption{Transaction execution in Ethereum and Solana.}
    \label{fig:ethereum_vs_solana_exe}
\end{figure}

For example, \Cref{fig:ethereum_vs_solana_exe} illustrates Ethereum's \rwOblivious{} and Solana's \rwAware{} execution using a block $B_i$ containing three transfer transactions. In Ethereum model (\Cref{fig:ethereum_vs_solana_exe}a), $T_1,\ T_2,\ \text{and},\ T_3$ are executed sequentially in order. The parallel execution of these transactions in the Ethereum VM (EVM) requires optimistic execution since access sets are not known a priori . 
Consequently, if transactions are executed optimistically in parallel, they must be validated and potentially rolled back in case of conflicts to satisfy the \presetSerialization{}. 
For example, if $T_3$ executes before $T_2$, it must be aborted and re-executed after $T_2$ commits (\TryC{$_2$}). In the Solana model, access sets are known in advance before execution, as shown in~\Cref{fig:ethereum_vs_solana_exe}b. This allows the runtime to detect that $T_2$ and $T_3$ conflict on $X_3$ and $X_4$, while $T_1$ is independent. Hence, 
can safely execute $T_1$ in parallel with $T_2$ and defer $T_3$, demonstrating how this model allows for efficient and conflict-aware parallel execution. 

\noindent\textbf{Research Goals:} We seek to understand the limits of parallelism in \rwOblivious{} and \rwAware{} execution models by conducting a comprehensive empirical study of the underlying conflict patterns. To our knowledge, no previous study has performed a comparative analysis of Ethereum and Solana, which represent distinct execution models. This leaves a critical gap in understanding how execution design choices shape conflict characteristics and thus the realizable throughput achievable through parallelism. Such an analysis allows us to answer the following questions:
\begin{description}
    \item[RQ1] Quantify what fraction of transactions are independent and how are transactions distributed between conflict families?
    \item[RQ2] What are typical depth and width characteristics of conflict chains in historical blocks?
    \item[RQ3] How does the maximum achievable parallelism compare between \rwAware{} and \rwOblivious{} execution models? 
    \item[RQ4] What are the fundamental design trade-offs in block transactional execution across different VMs, particularly the Ethereum VM (EVM) and the Solana VM (SVM)?
\end{description}

By analyzing historical traces, we aim to identify conflict families, independent transactions, and structural patterns across blocks. This will help to design conflict-aware scheduling, adaptive execution models, and virtual machine agnostic parallel execution that achieve higher throughput under realistic workloads. It also provides empirical insight into the frequency and granularity of conflicts, enabling us to assess the potential gains and limitations of parallelism.

\section{System Model}\label{sec:sm}
This section formalizes the execution semantics of the blockchain virtual machine (VM) from a block-execution perspective. 
The VM definition provides a foundation, while the VM-specific formalizations of \preset{} serializable execution highlight differences in state management.

\subsection{Blockchain Transaction Execution} 
A blockchain virtual machine is a deterministic state machine that processes \emph{transactions} and updates the global state according to predefined transition functions. Formally, it can be defined as a tuple $(\mathcal{S}, \mathcal{B}, \mathcal{E}, \mathcal{R})$, where $\mathcal{S}$ is the set of all possible blockchain states. $\mathcal{B}$ is the set of all possible blocks, each containing a list of \emph{transactions} $\mathcal{T}$. $\mathcal{E}: \mathcal{S}\times\mathcal{B}\to\mathcal{S}$ is the transition function that maps a state and a block to a new state. $\mathcal{R}: \mathcal{S} \times \mathcal{B} \to \{0,1\}$ is the validity function that determines whether a block is valid in a given state. 

A \emph{transaction} is a sequence of operations performed on a set of \emph{VM states}. 
For each transaction $T_k$, a VM must support the following operations: 
\Read$_k$($s$), where $s$ is an account or a state, that returns a value in
a domain $V$
or a special value $A_k\notin V$ (\emph{abort}),
\Write$_k$($X$, $v$), for a value $v \in V$,
that returns $\mathit{ok}$ or $A_k$, and a special final operation
$\mathit{\TryC}_k$ that returns $C_k\notin V$ (\emph{commit}) or $A_k$.
When $\mathit{\TryC}_k$ returns $C_k$, the transaction $T_k$ is deemed to have \emph{completed}, having completed its sequence of reads and writes.
For a transaction $T_k$, we denote all VM states accessed by its read and write as \readset{$_k$} (or \Rset{$_k$}) and \writeset{$_k$} (or \Wset{$_k$}), respectively. We denote by $\accessset{_k}$ (or $\Dset(T_k)$) the set of state locations accessed, read or written by $T_k$. {Execution of a block is the sequence of request–response events generated by these operations, where each operation consists of an atomic invocation (request) and corresponding completion (response).}

\noindent\textbf{Conflict:} Two transactions $T_i, T_j \in B$ are said to be in conflict if they both access a shared state $s \in \mathcal{S}$ and at least one of the operations is a write. Formally, a conflict occurs if:
\begin{equation}\label{eq:bvm_conf}
    \exists s \in \mathcal{S}, B = \{T_i, T_j, \dots\},\ \mathcal{E}(\mathcal{E}(s, T_i), T_j) \neq \mathcal{E}(\mathcal{E}(s, T_j), T_i)
\end{equation}

This condition implies that the execution order is crucial for the final state if the transactions are executed in parallel or reordered within the block $B$. Hence, the execution of $T_i$ and $T_j$ is order-dependent (on \preset{} order) within $B$ and are conflicting. 
 A historical analysis 
enables empirical evidence of conflict percentages and maximum parallelism that can be exploited in the real-world workloads.

\subsection{\Preset{} Serializable Executors} 
We formally define \emph{\preset{} serializability} as the correctness criterion for parallel execution in modern blockchains (adopting from \cite{Anjana2025BTMSSS}). Our goal is to capture a unified execution model that applies both to \textit{\RwOblivious{} runtimes} (EVM), and to \textit{\RwAware{} runtimes} (Solana's Sealevel~\cite{solanaSealevel}). 


\noindent\textbf{State Space.} Let $\mathcal{S}$ denote the universe of states (e.g., accounts and contract storage slots). A \emph{global state} is a mapping $\sigma: \mathcal{S}\to\mathcal{V}$ for some value domain $\mathcal{V}$. A block $B$ is a finite sequence of transactions $B = \{T_1, T_2, \dots, T_n\}$. Each transaction $T_i$ is a deterministic state transformer with an associated (static or dynamic) $\Rset_i \subseteq \mathcal{S}$ and $\Wset_i \subseteq \mathcal{S}$:
\[
  T_i : \Sigma \to \Sigma, \qquad
  T_i(\sigma) = \sigma'
\]
where: $\Sigma$ is the set of all global states, and the effect of $T_i$ may depend only on the values of $\sigma$ restricted to $\Rset_i$, while it may modify the values in $\Wset_i$.


\noindent\textbf{History.} A \emph{history} is the subsequence of an execution consisting of the invocation and response events of transaction operations. Two histories $H$ and $H'$ are \emph{equivalent} if \txns($H$) = \txns($H'$) and for every $T_k\ \in$\ \txns($H$),\ $H|k=H^\prime|k$.

\noindent
\noindent\textbf{Sequential History.} A history $H$ is said to be \emph{sequential} (denoted $H^{\mathsf{seq}}$) if the events of each transaction $T_k \in \txns(H)$ appear contiguously in $H$. A sequential history is obtained by executing transactions in a fixed order, called the \preset{} order, such that: $\forall\, T_i, T_j \in \txns(H^{\mathsf{seq}}),\ T_i\to T_j \Rightarrow \text{ All } E|i \text{ precede those of } E|j \text{ in } H^{\mathsf{seq}}.$

\noindent\textbf{\Preset{} Serializability.} A history $H$ for a block $B$ is an interleaving of the read-write operations of $T_1,\dots,T_n$ starting from some initial state $\sigma_0$, together with a final state $\sigma_H$.  Let $B^{\mathsf{seq}}$ denote the \emph{\preset{} serial execution} of the block, which executes \{$T_1,\dots, T_n$\} strictly in block order, without interleaving, starting from the same initial state $\sigma_0$, resulting in the final
state $\sigma_{B^{\mathsf{seq}}}$.

\noindent\textbf{\Preset{} Serializable Execution.} An execution $H$ of block $B$ is \emph{\preset{} serializable} if state results from $H$ is equivalent to the sequential execution ($H^{\mathsf{seq}}$) that executes \{$T_1,\dots,T_n$\} in $B$'s \preset{} order. It means, for a initial states $\sigma_0$, $\sigma_H = \sigma_{B^{\mathsf{seq}}}$. So, a blockchain executor is \emph{\preset{} serializable} if, for any history for a block $B$ it always results the same state as sequential execution.

\noindent\textbf{1. \RWO{} Execution.} A \preset{} serializable execution is \emph{\rwOblivious{}} if, prior to execution, it has no sound over-approximation of the read and write sets of individual transactions. Formally, for each $T_i$ it does not know any $(\widehat{\Rset}_i,\widehat{\Wset}_i)$ satisfying $\Rset_i \subseteq \widehat{\Rset}_i$ and $\Wset_i \subseteq \widehat{\Wset}_i$ for all possible inputs. Instead, read-write sets are discovered during execution as individual account or contract storage slot accesses occur. A \rwOblivious{} executor attempts to execute multiple transactions in parallel, using either:
\begin{itemize}[nosep,leftmargin=*]
  \item \emph{optimistic execution}, employing a \preset{} serializable STM-based executor that rolls-back and retries conflicting transactions~\cite{blockstm,iBTM_AFT_2025}; or
  \item \emph{lock-based execution}, where conflicts are detected during the acquisition of fine-grained locks on accounts or storage slots in the state space $\mathcal{S}$~\cite{saraph2019empirical}.
\end{itemize}

\begin{definition}
A blockchain executor is a \emph{\rwOblivious{} \preset{} serializable executor} if
  \begin{enumerate}[label=(\roman*)]
    \item it does not have any sound approximation (static or declared) of $(\Rset_i, \Wset_i)$ before executing $T_i$;
    \item it execute transaction in block $B$ arbitrarily in an interleaved manner, subject to a concurrency control mechanism that enforces that the resulting history $H$ is \preset{} serializable (i.e., produce the same state as results from the serial execution of $T_1,\dots,T_n$ in \preset{} order).
  \end{enumerate}
\end{definition}

\noindent
\textit{Ethereum as a \rwOblivious{} executor.} The standard Ethereum execution model is \rwOblivious{} where transactions do not declare their access sets, and any parallel executor must infer conflicts on-the-fly via fine-grained locking or optimistic execution while preserving equivalence with the block's \preset{} order.

\vspace{.8mm}
\noindent\textbf{2. \RWA{} Execution.} A \preset{} serializable execution is \emph{read--write aware} if, for each transaction $T_i$ in block $B$, it knows (before scheduling) declared read and write sets ($\widehat{\Rset}_i, \widehat{\Wset}_i \subseteq \mathcal{S}$) such that for all concrete executions $\Rset_i \subseteq \widehat{\Rset}_i \text{ and } \Wset_i \subseteq \widehat{\Wset}_i$. We call $(\widehat{\Rset}_i,\widehat{\Wset}_i)$ the \emph{declared access sets}. The executor uses only these declared sets to decide which transactions may run in parallel. It must ensure that no two concurrently executed transactions have a potential conflict according to the declared sets, i.e.,
\vspace{-.1cm}
\begin{equation}
  \label{eq:\rwAware{}-safety}
  \begin{aligned}
  \forall i \neq j:
  \bigl(
    (\widehat{\Wset}_i \cap \widehat{\Wset}_j) \cup
    (\widehat{\Wset}_i \cap \widehat{\Rset}_j) \cup\\\quad
    (\widehat{\Rset}_i \cap \widehat{\Wset}_j)
  \bigr) = \emptyset \\\quad
  \quad\Rightarrow\quad
  T_i \text{ and } T_j \text{ may be executed in parallel.}
  \end{aligned}
\end{equation}

\begin{definition}
A blockchain executor is an \emph{\rwAware{} \preset{} serializable  executor} if
  \begin{enumerate}[label=(\roman*)]
    \item it results only histories that are \preset{} serializable with respect to the block transaction order; and
    \item its parallelization decisions are made exclusively using declared access sets $(\widehat{\Rset}_i,\widehat{\Wset}_i)$ satisfying $\Rset_i \subseteq \widehat{\Rset}_i$ and $\Wset_i \subseteq \widehat{\Wset}_i$.
  \end{enumerate}
\end{definition}

\noindent
\textit{Solana as a \rwAware{} executor.} The Solana runtime is an instance of a \rwAware{} \preset{} serializable executor where the account lists and their read-write modes play the role of declared access sets $(\widehat{\Rset}_i,\widehat{\Wset}_i)$ with transactions, and the scheduler forbids concurrent execution of transactions whose declared sets conflict.

To summarize, Solana's account-based execution is a \rwAware{} \preset{} serializable executor, whereas any exiting parallel EVM executors~\cite{pevm,seiparallelexe,polygonParallelization} operate in the stricter \rwOblivious{} setting unless equipped with an external conflict analyzer~\cite{iBTM_AFT_2025} or access-list mechanism~\cite{feist2025block} that lifts it into the \rwAware{} model by sound over-approximations $(\widehat{\Rset}_i,\widehat{\Wset}_i)$ for each transaction. In our historical block analysis, we respect the \preset{} transaction order to ensure that, if parallel execution is implemented on mainnet by validators, it results in the same state as the \preset{} serialization order.

\section{Defining and Extracting Conflict Metrics}\label{sec:conf_analysis}
In this section, we first introduce conflict metrics 
to evaluate parallelism in historical blocks. Then we discuss the historical block data extractions for the analysis. 




\subsection{Conflict Metrics}
Let's consider the following set of transactions \{$T_1, T_2, \dots, T_8$\} to discuss the {conflict metrics}. For simplicity, we consider small read-write sets, although in practice these sets can be significantly larger. 

\vspace{.05cm}
\noindent
\fbox{\begin{minipage}{1\linewidth}
\vspace{-.2cm}
\footnotesize
\[
\begin{alignedat}{2}
T_1\!: &~\texttt{transfer(X$_1$, X$_2$, \$5)},      \quad&~\Rset_1 = \{X_1\},\  &~\Wset_1 = \{X_1, X_2\} \\
T_2\!: &~\texttt{transfer(X$_3$, X$_4$, \$7)},      \quad&~\Rset_2 = \{X_3\},\  &~\Wset_2 = \{X_3, X_4\} \\
T_3\!: &~\texttt{transfer(X$_3$, X$_6$, \$10)},     \quad&~\Rset_3 = \{X_3\},\  &~\Wset_3 = \{X_3, X_6\} \\
T_4\!: &~\texttt{transfer(X$_6$, X$_7$, \$5)},      \quad&~\Rset_4 = \{X_6\},\  &~\Wset_4 = \{X_6, X_7\} \\
T_5\!: &~\texttt{transfer(X$_8$, X$_9$, \$2)},      \quad&~\Rset_5 = \{X_8\},\  &~\Wset_5 = \{X_8, X_9\} \\
T_6\!: &~\texttt{transfer(X$_9$, X$_{10}$, \$1)},   \quad&~\Rset_6 = \{X_9\},\  &~\Wset_6 = \{X_9, X_{10}\} \\
T_7\!: &~\texttt{get\_balance(X$_{11}$)},           \quad&~\Rset_7 = \{X_{11}\}, &~\Wset_7 = \{\} \\
T_8\!: &~\texttt{get\_balance(X$_{11}$)},           \quad&~\Rset_8 = \{X_{11}\}, &~\Wset_8 = \{\}
\end{alignedat}
\]
\end{minipage}}

   \vspace{5pt}
    \noindent\textbf{1. Conflict dependent transactions.} Transactions $T_i$ and $T_j$ are said to be \textbf{conflict dependent} if: 
    $T_i$ precedes $T_j$ in the \emph{\preset{}} order and $T_i$ and $T_j$ in conflict.
        \begin{equation}
        \text{Conflict}(T_i, T_j, s) =
            \begin{cases}
                1, & \text{if } 
                    s \in \Wset_i \cap \Wset_j \;\lor\;\\ & \quad 
                    s \in \Wset_i \cap \Rset_j \;\lor\; \\ & \quad 
                    s \in \Rset_i \cap \Wset_j \\
                0, & \text{otherwise.}
            \end{cases}
        \end{equation}
         
         It means that the read-write sets of $T_i$ and $T_j$ have a non-empty access on at least one conflicting state.
         
        \emph{Example:} $T_2$ and $T_3$ are in \emph{write-read}: \Wset{$_2$} $\cap$ \Rset{$_3$} = \{$X_3$\}, and \emph{write-write}: \Wset{$_2$} $\cap$ \Wset{$_3$} = \{$X_3$\} conflicts, hence $T_3$ is conflict dependent on $T_2$. Similarity $T_3-T_4$ and $T_5-T_6$ are conflict dependent. 

    
    \vspace{.05cm}
    \noindent\textbf{2. Conflict independent transactions.} The number of independent transactions in the block measures how much of the workload can be executed in parallel. A higher percentage of independent transactions suggests that the blockchain can scale better with more cores. Let B = $\{T_1, T_2, \ldots, T_n\}$ be the set of transactions in a block. Let $I \subseteq B$ be the set of \textbf{independent} transactions such that $\forall T_i \in I, \forall T_j \in B, i \neq j: T_i \text{ and } T_j \text{ do not conflict.}$ Then the \textbf{percentage of independent transaction} is defined as: $\frac{|I|}{|B|} \times 100\%$.
    
    \emph{Example:} $T_1$, $T_7$, and $T_8$ are independent transactions, since read-read is not a conflicting operation and there is no overlap with other writing transactions. The independent percentage in our example set of transactions is $\frac{3}{8} \times 100\% = 37.5\%$. 
    
    
    \vspace{.05cm}
    
    \noindent\textbf{3. Longest conflict chain (LCC).} The conflict chain or critical path~\cite{BlumofeCilk,BlumofeWorkStealing,HM93,GraceBerger,Anjana2025BTMSSS}, 
    represents the conflicting sequence where each transaction conflicts with the next in the \preset{} order. It quantifies minimal execution time under parallelism and bounds the maximum speed-up based on the length of sequential dependencies~\cite{Anjana2025BTMSSS}. Even with the number of available processors, the algorithm must execute the chain transactions sequentially. 
    The longer the chain, the lower the speedup. 
    
     To compute the longest chain of conflicting transactions, let's construct a conflict graph $G = (V, E)$, where each vertex $v_i \in V$ represents a transaction $T_i$, and a directed edge $(v_i, v_j) \in E$ if $T_i$ and $T_j$ conflict. Then, the longest conflict chain is:
    \vspace{-.15cm}
    \begin{equation*}
       \text{LCC} = \max_{P \in \text{Paths}(G)} |P|
    \end{equation*}
    \emph{Example:} There are two conflict chains:
    $c_1$:  \circled{\small$T_2$} $\to$ \circled{\small$T_3$} $\to$ \circled{\small$T_4$} and 
    $c_2$:  \circled{\small$T_5$} $\to$ \circled{\small$T_6$}. 
    The longest chain is $c_1$, since $|c_1|$ $>$ $|c_2|$.
    
    
    \vspace{.05cm}
   \noindent\textbf{4. Conflict families (CF).} A \textit{conflict family} can be viewed as a {connected component} in the conflict graph $G$. That is, a set of transactions $F \subseteq B$ s.t. $\forall T_i,\ T_j \in F,\ \exists \text{ a path in } G \text{ between } v_i - v_j$ and \emph{$F$ is maximal}, i.e., no larger set than $F$. Formally, $F_1, F_2, F_3, \dots$ are all connected components of $G$, i.e., $\{F_1, F_2, \ldots, F_k\}$ denote disjoint conflict families.
    \vspace{-.15cm}
    \begin{equation*}
        B = \bigcup_{i=1}^k F_i, \quad F_i \cap F_j = \emptyset \text{ for } i \neq j 
    \end{equation*}
    \vspace{-.15cm}
    \emph{Example:} The following are the conflict families in the transaction set of our example:
    \vspace{-.2cm}
    \begin{multicols}{2}
        \begin{itemize}
            \item $F_1$: $\{T_1\}$
            \item $F_2$: $\{T_2, T_3, T_4\}$
            \item $F_3$: $\{T_5, T_6\}$
            \item $F_4$: $\{T_7\}$
            \item $F_5$: $\{T_8\}$
        \end{itemize}
    \end{multicols}
    \vspace{-.2cm}
    
    There are five conflict families; although both $T_7$ and $T_8$ read $X_{11}$ they belong to separate conflict families, since reads are non-conflicting operations. 

    
    \vspace{.05cm}
    \noindent\textbf{5. Most dense conflict family (MDCF)} 
    is the conflict family with most conflicting transactions: $F^* = \arg\max_{F_i} |F_i|$. It sets a theoretical bound on execution speedup, the size and density of the largest conflict family limit the achievable parallelism: {more dense} $\rightarrow$ {more serialization} $\rightarrow$ {less parallelism}. 
    In our example, the MDCF is $F_2$ with 3 transactions.
    
    
    \vspace{.05cm}
    \noindent\textbf{6. Total and Write-write conflicts.} The \textbf{total conflicts (TC)} measure how many transaction pairs access overlapping state in conflicting ways (r-w or w-w), indicating the general level of transaction interdependency. Higher total conflicts mean higher contention, suggesting lower parallelism without advanced conflict management. On the other hand, {write-write conflicts} are stronger conflicts 
    that always require serialization. However, multi-versioning allows transactions to proceed in parallel by creating separate versions of conflicting writes, thereby increasing parallelism even in the presence of w-w conflicts. In our analysis, total conflicts are determined by individual state accesses among transactions in \preset{} order, where each pair of conflicting accesses contributes one conflict. 
    The \textbf{write-write conflicts} (w-w conflicts) are at the level of individual states written by the transactions. 
    There are $6$ total conflicts in our example and $3$ w-w conflicts as follows:
    \begin{itemize}
        \item \circled{\small$T_2$} $\to$ \circled{\small$T_3$}: \Write{}-\Read{} and \Write{}-\Write{} conflict on $X_3$.
        \item \circled{\small$T_3$} $\to$ \circled{\small$T_4$}: \Write{}-\Read{} and \Write{}-\Write{} conflict on $X_6$.
        \item \circled{\small$T_5$} $\to$ \circled{\small$T_6$}: \Write{}-\Read{} and \Write{}-\Write{} conflict on $X_9$.
    \end{itemize}


\vspace{.25cm}
For these conflict metrics, we analyze transactions from different \textbf{historical periods} (HPs)-- defined as contiguous time windows 
(e.g., grouped by block ranges or timestamps) for which blocks are analyzed. Different HPs exhibit varying characteristics in terms of transaction load and state access patterns. Therefore, to capture the temporal dynamics of transaction conflict, we performed our analysis over multiple HPs on both blockchain networks.

\subsection{Extracting Historical Data} 
\paragraph{Ethereum} We segregate Ethereum transactions into two types: ETH transfer (or simply transfer) and smart contract transactions (SCTs), to analyze 
conflict patterns. The transfer transactions perform pure value transfers between externally owned addresses (EOAs) or to smart contract addresses. In contrast, SCTs interact with sender addresses and contract address(es) to modify blockchain state via function calls and storage updates within the contract(s).

We reconstruct fine-grained read--write sets (r-w sets) for each transaction in the block using Ethereum’s low-level execution tracing. We invoke \texttt{debug\_traceBlockByNumber} with the \texttt{prestateTracer}~\cite{chainstackTraceblock} in both \texttt{diff=false} and \texttt{diff=true} modes to get the block pre-state and the exact state mutations induced by execution. Comparing accessed pre-states with post-execution diffs allows us to extract account- and storage-slot-level reads and writes, which could be EOAs, contract addresses, and storage slots with in contracts. This enables the detection of potential conflicts (r-w, w-w, and w-r) arising from overlapping state modifications by transactions enforced by EVM semantics, thereby forming the basis of our historical analysis.


It should be noted that transactions are independent if they are initiated by separate EOAs and access different addresses and storage locations within all the invoked contracts. However, there is a special case in which every transaction updates the Coinbase account (the block proposer’s account) for fee payment. As a result, all transactions are logically in conflict, unless the Coinbase account is treated as an exception. For this reason, when we analyze conflicts, we remove the Coinbase account from the transaction \accessset{} (or \Dset{}). In~\cite{pevm}, a solution is proposed to collect the fee payment for each transaction independently, and to update the Coinbase account cautiously at the end of the block.


\paragraph{Solana} Solana transactions are made up of the account-access specification, a list of accounts to be read (\readset{}) or written (\writeset{}). 
This specification is added to the transactions upfront by the clients through RPC interaction. The success or failure of a transaction depends on the freshness of its r-w sets, from the moment it is added by the client until its execution at the validator node. 
The \readset{} and \writeset{} simplify our 
conflict detection and analysis for Solana blocks. We used the beta API of Solana's mainnet to obtain block details in JSON format with the max-supported transaction version set to 0~\cite{solanaGetblock}. The extracted details are then parsed to obtain the information required for our analysis.
\section{Ethereum's Empirical Study\label{sec:eth_analysis}}
As shown in \Cref{tab:ethereumHP}, we selected three distinct historical periods (HPs) based on the major events that impacted Ethereum's network congestion. Each of these periods allows us to analyze blocks and gain insights 
into how major events, such as popular \dApp{} launches and significant protocol optimizations, affect transaction parallelism and conflicts. This also helps us understand the limitations of parallel execution approaches under different network conditions. 

\begin{table}[!tb]
    \caption{Historical blocks from Ethereum's mainnet
    }
    \label{tab:ethereumHP}
    \resizebox{1\columnwidth}{!}{%
        \begin{tabular}{|r|c|c|c|}
        \hline
        & \textbf{\begin{tabular}[c]{@{}c@{}}~~CryptoKitties~~~~~~\\Deployment (E$_{ck}$)\end{tabular}} 
        & \textbf{\begin{tabular}[c]{@{}c@{}}~~~~Ethereum~2.0~~~~~\\Merge (E$_{2o}$)\end{tabular}} 
        & \textbf{\begin{tabular}[c]{@{}c@{}}~~~~Ethereum Recent~~~~\\Blocks (E$_{rb}$)\end{tabular}} 
        \\ \hline\hline
            \textbf{Historical Event Block} 
            & 4605167          
            & 15537393            
            & 21631500            
            \\ 
                
            \textbf{Blocks Before Event}  
            & 4604664 - 4605166   
            & 15536879 - 15537392   
            & 21631000 - 21631500 
            \\ 
            
            \textbf{Blocks After Event}   
            & 4605168 - 4605670 
            & 15537394 - 15537907 
            & 21631501 - 21632001         
            \\ \hline 
            
        \end{tabular}%
    }
\vspace{-.45cm}
\end{table}
\noindent\textbf{CryptoKitties Contract Deployment ($E_{ck}$).} The CryptoKitties~\cite{cryptokitties} game is among the first and most popular \dApp{s}. The contract was deployed in block 4605167, after which an unexpected surge in transactions led to high congestion on the Ethereum network.. The $E_{ck}$ workload consists of 500 blocks each, before (E$_{ck-before}$) and after (E$_{ck-after}$) the contract deployment. This period is characterized by a high volume of transactions concentrated on a single contract, resulting in significant contract-level congestion, as also observed in~\cite{saraph2019empirical}. Consequently, this workload serves as a representative benchmark for evaluating how effectively parallel execution approaches handle large influxes of transactions targeting the same contract.

\noindent\textbf{Ethereum~2.0 Merge ($E_{2o}$).} Ethereum's transition from proof-of-work to proof-of-stake consensus took place in block 15537393, called the Ethereum~2.0 Merge~\cite{ethereum2merge}. This event changed consensus and optimized transaction processing, including block validation. Hence, in this workload, we analyze benefits of this upgrade on parallel execution, throughput, and network traffic by comparing 500 blocks before and 500 after the merge.

\noindent\textbf{Recent Blocks ($E_{rb}$).} We also analyze 1000 recent blocks, ranging from 21631000 to 21632001. We selected this range to better understand current transaction access patterns and parallelism under normal conditions, where no major historical events impacting network traffic. We compare different historical periods by tracking access patterns, network congestion over time, and the impact of optimizations on recent blocks. This analysis also helps develop approaches that take advantage of parallelism efficiently for future network upgrades.

\subsection{Analysis}
As illustrated in \Cref{fig:eth_hp_analysis,fig:ethereum-hp,fig:rw-family-cluster,fig:rw-speedup-vs-chain}, and \Cref{tab:ethereum_rb}, the following observations can be made for real-world Ethereum data:

\begin{figure}[!tb]
    \scalebox{1.2}{\begin{tikzpicture}[scale = 1]
    \begin{axis}[
        base,
        nnc,
        ybar   = .4pt,
        xtick  = data,
        ymin   = 0,
        ymax   = 360,
        ytick  = {0, 50, 100, 150, 200},
        width  = 16cm, 
        height = 10cm,
        enlarge x limits = 0.06,
        ylabel           = {Transaction Count},
        y label style	 = {at = {(-.11,.39)}},
        xticklabel style = {rotate=35, anchor=east},
        bar width         = 5.75pt,
        nodes near coords,
        every node near coord/.append style={font = \tiny, rotate = 90, xshift = 7pt, yshift = -5pt},
        legend style = {at = {(1.14,.86)}, anchor = north, legend columns = 2, draw = none, font = \tiny},
        symbolic x coords = {Block Size, ETH Transfer, Smart Contract Txs, ERC20 Transfer, Independent Txs, Longest Chain of Conflicts, Independent ETH Txs, Independent SC Txs, Conflict Families},
        xticklabels = {Size, ETH, SCTs, ERC20, Ind, LCC, IETH, ISCTs, CF},
    ]
    
        \addplot[ybar, black!80, fill=black!80, every node near coord/.append style={xshift = 0pt, yshift = -5pt}] 
        coordinates {
            (Block Size, 71) 
            (ETH Transfer, 35) 
            (Smart Contract Txs, 37) 
            (ERC20 Transfer, 12) 
            (Independent Txs, 38) 
            (Longest Chain of Conflicts, 15) 
            (Independent ETH Txs, 21) 
            (Independent SC Txs, 17) 
            (Conflict Families, 30)
        };

        \addplot[ybar, black!25, fill=black!25, every node near coord/.append style={xshift = 0pt, yshift = -3pt}] 
        coordinates {
            (Block Size, 83) 
            (ETH Transfer, 42) 
            (Smart Contract Txs, 41) 
            (ERC20 Transfer, 13) 
            (Independent Txs, 42) 
            (Longest Chain of Conflicts, 15) 
            (Independent ETH Txs, 24) 
            (Independent SC Txs, 19) 
            (Conflict Families, 37)
        };

        \addplot[ybar, red, fill=red, every node near coord/.append style={xshift = 0pt, yshift = -1pt}] 
        coordinates {
            (Block Size, 178) 
            (ETH Transfer, 66) 
            (Smart Contract Txs, 113) 
            (ERC20 Transfer, 11) 
            (Independent Txs, 87) 
            (Longest Chain of Conflicts, 38) 
            (Independent ETH Txs, 32) 
            (Independent SC Txs, 56) 
            (Conflict Families, 56)
        };

        \addplot[ybar, red!50, fill=red!50, every node near coord/.append style={xshift = 0pt, yshift = 1pt}]
        coordinates {
            (Block Size, 156) 
            (ETH Transfer, 42) 
            (Smart Contract Txs, 114) 
            (ERC20 Transfer, 16) 
            (Independent Txs, 55) 
            (Longest Chain of Conflicts, 42) 
            (Independent ETH Txs, 24) 
            (Independent SC Txs, 32) 
            (Conflict Families, 40)
        };

        \addplot[ybar, teal!90, fill=teal!90, every node near coord/.append style={xshift = 0pt, yshift = 3pt}]
        coordinates {
            (Block Size, 181) 
            (ETH Transfer, 65) 
            (Smart Contract Txs, 116) 
            (ERC20 Transfer, 43) 
            (Independent Txs, 92) 
            (Longest Chain of Conflicts, 31) 
            (Independent ETH Txs, 49) 
            (Independent SC Txs, 45) 
            (Conflict Families, 72)
        };

        \addplot[ybar, teal!40, fill=teal!40, every node near coord/.append style={xshift = 0pt, yshift = 5pt}] 
        coordinates {
            (Block Size, 176) 
            (ETH Transfer, 62) 
            (Smart Contract Txs, 114) 
            (ERC20 Transfer, 38) 
            (Independent Txs, 90) 
            (Longest Chain of Conflicts, 27) 
            (Independent ETH Txs, 47) 
            (Independent SC Txs, 45) 
            (Conflict Families, 69)
        };
        
        \legend{~E$_{ck-before}$~~~~, ~E$_{ck-after}$, ~E$_{e2-before}$~~~~, ~E$_{2o-after}$, ~E$_{rb-before}$~~~~, ~E$_{rb-after}$}

        \node[anchor=north west] at (axis cs: Block Size, 367) {
            \hspace{-.8cm}
            \resizebox{1.18\columnwidth}{!}{%
                \begin{tabular}{|r|c|c|c|}
                \hline
                & \textbf{\begin{tabular}[c]{@{}c@{}}CryptoKitties (E$_{ck}$)\end{tabular}} 
                & \textbf{\begin{tabular}[c]{@{}c@{}}Ethereum~2.0 Merge (E$_{2o}$)\end{tabular}} 
                & \textbf{\begin{tabular}[c]{@{}c@{}}Recent Blocks (E$_{rb}$)\end{tabular}} 
                \\ 
                      
                    & {\textcolor{black!80}{before event} - \textcolor{black!25}{after event~~}}
                    & {\textcolor{red}{before event} - \textcolor{red!50}{after event~~}}
                    & {\textcolor{teal!90}{before event} - \textcolor{teal!40}{after event~~}}           
                    \\ \hline
                    
                    \textbf{Block Size}       
                    & 71 - 83  
                    & 178 - 156 
                    & 181 - 176
                    \\
        
                    \textbf{ETH Transfer Transactions (ETH)}       
                    & 35 - 42   
                    & 66 - 42 
                    & 65 - 62
                    \\
        
                    \textbf{Smart Contract Transactions (SCTs)}       
                    & 37 - 41   
                    & 113 - 114
                    & 116 - 114
                    \\
        
                    \textbf{ERC20 Transfer Transactions (ERC20)}       
                    & 12 - 13 
                    & 11 - 16
                    & 43 - 38
                    \\
                    
                    \textbf{Independent Transactions (Ind) (\%)}       
                    & 35 (56.78\%) - 38 (49.77\%)  
                    & 95 (57.47\%) - 62 (44.53\%)
                    & 115 (64.48\%) - 111 (63.80\%)
                    \\
                    
                    \textbf{Longest Chain of Conflicts (LCC) (\%)}       
                    & 14 (23.09\%) - 14 (21.24\%)  
                    & 36 (19.43\%) - 36 (22.92\%)
                    & 14 (7.58 \%) - 15 (9.06\%)
                    \\
                    
                    \textbf{Independent ETH Transfer (IETH) (\%)}       
                    & 16 (58.10\%) - 18 (50.18\%)
                    & 28 (55.19\%) - 21 (61.22\%)
                    & 41 (66.11\%) - 39 (66.18\%)
                    \\
                    
                    \textbf{Independent SC Transactions (ISCTs) (\%)}       
                    & 18 (57.40\%) - 20 (53.81\%)  
                    & 68 (66.38\%) - 45 (43.62\%)
                    & 74 (65.30\%) - 72 (64.74\%)
                    \\
                    
                    \textbf{Conflict Families (CFs)}       
                    & 43 - 48    
                    & 108 - 78 
                    & 133 - 128
                    \\
                    
                    \textbf{Dense Conflict Family Size (DCFS)}       
                    & 14 - 14
                    & 36 - 35
                    & 14 - 15
                    
                    
                    \\\hline
                \end{tabular}%
            }
        };
    \end{axis}
\end{tikzpicture}}
    \caption{Etherem's mainnet: analysis of 1000 blocks per historical period using transaction read–write sets.}
    \label{fig:eth_hp_analysis}
    \vspace{-.35cm}
\end{figure}


\noindent\textbf{Observation-1.} The initial evaluation aims to characterize the parallelism by distinguishing between dependent (conflicting) and independent transactions, identifying the longest chain of conflicting transactions, and examining conflict families both within and across historical periods. As shown in~\Cref{fig:eth_hp_analysis}, transactions per block have increased since $E_{ck}$, with contract transactions increasing by more than 4$\times$ and ETH transfers (or transfers) by nearly 2$\times$. This implies increased demand for computational resources and broader adoption of blockchain for diverse applications (\dApp{s}). The percentage of independent transactions has also increased, and over 60\% are independent on average in recent blocks. The conflict chain comprises 21-23\% of the block size in $E_{ck}$ and $E_{2o}$, reaching a maximum of $\sim$23\% in $E_{ck}$ and decreasing to $\sim$7–9\% in recent blocks. This suggests that even with perfect parallelization of other transactions, including scheduling of conflicting transactions, the maximum speedup is restricted to $\sim$10\% of transactions that must be executed sequentially, the theoretical upper bound on speedup over sequential execution in recent blocks. 

Compared to previous HPs, the percentage of independent transfer and contract transactions in recent blocks has increased, 
indicating an upsurge in contract applications and more diverse user patterns for transfers. Further, the increase in conflict families and block sizes from the earlier period to the more recent one suggests that there is more room for parallel execution that can improve the throughput.


\noindent\textbf{Observation-2.} Calculating the ratio of transfers to contract transactions (table in \Cref{fig:eth_hp_analysis}) shows an increased user engagement with contract applications. In $E_{ck}$, the ratio of $\frac{ETH\ transfer}{SC\ Txs}$ is $\frac{38.5}{39}\approx0.99$, while it is $\sim$0.47 in $E_{2o}$ and $\sim$0.55 in $E_{rb}$. This indicates a surge in computational costs over time and highlights the need for parallel execution.

\noindent\textbf{Observation-3.} As shown in \Cref{fig:ethereum-hp}, we analyzed 1000 blocks from each HP to understand how many blocks in each HP have a certain percentage ($>$40\%, $>$50\%, $\cdots$, $>$80\%) of independent transactions, and which HP exhibits higher parallelism compared to the others. The number of blocks with over 40\% independent transactions has increased in recent blocks, and over 98\% of blocks have at least 40\% independent transactions in $E_{rb}$. However, there was a notable decline in independent transactions after each historical event, suggesting a shorter spike in conflicts. Note that over 50\% blocks in each HP had more than 50\% of independent transactions, while a significant drop is observed after after $E_{2o}$. The reasons could be high congestion for a specific contract (the conflict chain increased), a decrease in the number of transfer transactions, and a slight decrease in block size compared to pre-merge, as observed in \Cref{fig:ethereum-hp}. 

\begin{figure}[!tb]
    \scalebox{1.25}{\begin{tikzpicture}[scale = 1]
    \begin{axis}[
        base,
        nnc,
        ymax	= 575,
        ymin	= 0,
        xlabel	= {Percentage of Independent Transactions in the Block},
        xtick	= {data},
        width   = 15cm, 
        height  = 8cm,
        xticklabels={$>$40\%, $>$50\%, $>$60\%, $>$70\%, $>$80\%},
        ytick	= {10,100,200,300,400,500,600},
        ylabel 	= {Block Count},
        bar width = 10.25pt,
        y label style = {at = {(-.15,.65)}},
        ylabel style  = {align = center},
        title style   = {at = {(0.74,.85)}, font = \large, gray!50},
        legend style  = {at = {(1.15,1.15)}, anchor = south, legend columns = 2, fill = none, font = \tiny, draw = none}, 
    ]
    \addplot[
        ybar, 
        black!80,
        fill = black!80, 
        every node near coord/.append style = {
                xshift = 9pt, 
                yshift = 17pt, 
                rotate = 90, 
                anchor = east, 
            }
        ] 
        table[x index = 0, y index = 1, col sep = space]
        {figs/results/ethereum-before.txt};

    \addplot[
        ybar, 
        black!25,
        fill = black!25, 
        every node near coord/.append style = {
                xshift = 6pt, 
                yshift = 17pt, 
                rotate = 90, 
                anchor = east, 
            }
        ] 
        table[x index = 0, y index = 1, col sep = space]
        {figs/results/ethereum-after.txt};
    
    \addplot[
        ybar, 
        red!90,
        fill = red!90, 
        every node near coord/.append style = {
                xshift = 2pt, 
                yshift = 17pt, 
                rotate = 90, 
                anchor = east, 
            }
        ] 
        table[x index = 0, y index = 2, col sep = space]
        {figs/results/ethereum-before.txt};

    \addplot[
        ybar, 
        red!50,
        fill = red!50, 
        every node near coord/.append style = {
                xshift = -2pt, 
                yshift = 17pt, 
                rotate = 90, 
                anchor = east, 
            }
        ] 
        table[x index = 0, y index = 2, col sep = space]
        {figs/results/ethereum-after.txt};
    
    \addplot[
        ybar, 
        teal!90,
        fill = teal!90, 
        every node near coord/.append style = {
                xshift = -7pt, 
                yshift = 17pt, 
                rotate = 90, 
                anchor = east, 
            }
        ] 
        table[x index = 0, y index = 3, col sep = space]
        {figs/results/ethereum-before.txt};

    \addplot[
        ybar, 
        teal!40,
        fill = teal!40, 
        every node near coord/.append style = {
                xshift = -9pt, 
                yshift = 17pt, 
                rotate = 90, 
                anchor = east, 
            }
        ] 
        table[x index = 0, y index = 3, col sep = space]
        {figs/results/ethereum-after.txt};

    \legend{~E$_{ck-before}$~~~,~E$_{ck-after}$~~~,~E$_{2o-before}$~~~,~E$_{2o-after}$~~~,~E$_{rb-before}$~~~,~E$_{rb-after}$~~~}
    \end{axis}
\end{tikzpicture}}
    \caption{Ethereum: the number of blocks in which the percentage of independent transactions exceeds a threshold before and after a historical event.}
    \label{fig:ethereum-hp}
    \vspace{-.5cm}
\end{figure}

\begin{table*}
    \centering
    \caption{
    Block-wise conflict metrics of recent Ethereum blocks (21631001--21631020). The table shows transaction composition, independence ratios, and conflict structure in the \preset{} order. With an average block size of 176, a large fraction of transactions remain independent (66.17\% on average), while the conflict chain stays relatively small (6.92\% of block size). Several blocks show high conflict counts and dense conflict families (e.g., block~21631010). 
    }
    \label{tab:ethereum_rb}
    \resizebox{\textwidth}{!}{%
    \begin{tabular}{|c|c|c|c|c|c|c|c|c|c|c|c|c|c|}
        \hline
        \textbf{Block ID} &
        \textbf{\begin{tabular}[c]{@{}c@{}} Block Size \end{tabular}} &
        \textbf{\begin{tabular}[c]{@{}c@{}} ETH Transfer\\Txs \end{tabular}} &
        \textbf{\begin{tabular}[c]{@{}c@{}} SC Txs \end{tabular}} &
        \textbf{\begin{tabular}[c]{@{}c@{}} ERC20\\Transfer Txs \end{tabular}} &
        \textbf{\begin{tabular}[c]{@{}c@{}} Independent\\Txns (\%) \end{tabular}} &
        \textbf{\begin{tabular}[c]{@{}c@{}} Chain of\\Conflicts (\%) \end{tabular}} &
        \textbf{\begin{tabular}[c]{@{}c@{}} Independent ETH\\Transfer Txs (\%) \end{tabular}} &
        \textbf{\begin{tabular}[c]{@{}c@{}} Independent\\SC Txs (\%) \end{tabular}} &
        \textbf{\begin{tabular}[c]{@{}c@{}} Conflict\\Families \end{tabular}} &
        \textbf{\begin{tabular}[c]{@{}c@{}} Dense Conflict\\Family \end{tabular}} &
        \textbf{\begin{tabular}[c]{@{}c@{}} Total\\Conflicts \end{tabular}}&
        \textbf{\begin{tabular}[c]{@{}c@{}} Write-Write\\Conflicts \end{tabular}}
        \\ \hline\hline

        \textbf{\href{https://etherscan.io/block/21631001}{21631001}} &
        337 &
        130 &
        207 &
        75 &
        186 (55.19\%) &
        17 (5.04\%) &
        69 (53.08\%) &
        120 (57.97\%) &
        223 &
        17 &
        453 &
        439
        \\

        \textbf{\href{https://etherscan.io/block/21631002}{21631002}} &
        148 &
        39 &
        109 &
        46 &
        91 (61.49\%) &
        9 (6.08\%) &
        27 (69.23\%) &
        64 (58.72\%) &
        108 &
        9 &
        83 &
        80
        \\
        
        \textbf{\href{https://etherscan.io/block/21631003}{21631003}} &
        82 &
        27 &
        55 &
        35 &
        57 (69.51\%) &
        8 (9.76\%) &
        18 (66.67\%) &
        39 (70.91\%) &
        65 &
        8 &
        40 &
        40
        \\
        
        \textbf{\href{https://etherscan.io/block/21631004}{21631004}} &
        191 &
        65 &
        126 &
        59 &
        127 (66.49\%) &
        11 (5.76\%) &
        44 (67.69\%) &
        83 (65.87\%) &
        147 &
        11 &
        97 &
        92
        \\
        
        \textbf{\href{https://etherscan.io/block/21631005}{21631005}} &
        233 &
        75 &
        158 &
        55 &
        152 (65.24\%) &
        15 (6.44\%) &
        61 (81.33\%) &
        95 (60.13\%) &
        178 &
        15 &
        86 &
        73
        \\
        
        \textbf{\href{https://etherscan.io/block/21631006}{21631006}} &
        154 &
        64 &
        90 &
        34 &
        110 (71.43\%) &
        11 (7.14\%) &
        42 (65.63\%) &
        69 (76.67\%) &
        124 &
        11 &
        94 &
        87
        \\
        
        \textbf{\href{https://etherscan.io/block/21631007}{21631007}} &
        192 &
        67 &
        125 &
        46 &
        132 (68.75\%) &
        8 (4.17\%) &
        48 (71.64\%) &
        85 (68.00\%) &
        152 &
        8 &
        70 &
        68
        \\
        
        \textbf{\href{https://etherscan.io/block/21631008}{21631008}} &
        163 &
        60 &
        103 &
        47 &
        97 (59.51\%) &
        10 (6.13\%) &
        39 (65.00\%) &
        60 (58.25\%) &
        121 &
        10 &
        89 &
        97
        \\
        
        \textbf{\href{https://etherscan.io/block/21631009}{21631009}} &
        177 &
        68 &
        109 &
        55 &
        107 (60.45\%) &
        26 (14.69\%) &
        47 (69.12\%) &
        61 (55.96\%) &
        123 &
        26 &
        374 &
        362
        \\
        
        \textbf{\textcolor{red}{\href{https://etherscan.io/block/21631010}{21631010}}} &
        \textcolor{red}{207} &
        \textcolor{red}{86} &
        \textcolor{red}{121} &
        \textcolor{red}{52} &
        \textcolor{red}{107 (51.69\%)} &
        \textcolor{red}{34 (16.43\%)} &
        \textcolor{red}{30 (34.88\%)} &
        \textcolor{red}{77 (63.64\%)} &
        \textcolor{red}{125} &
        \textcolor{red}{34} &
        \textcolor{red}{832} &
        \textcolor{red}{826} 
        \\
        
        \textbf{\href{https://etherscan.io/block/21631011}{21631011}} &
        148 &
        46 &
        102 &
        31 &
        97 (65.54\%) &
        13 (8.78\%) &
        35 (76.09\%) &
        62 (60.78\%) &
        112 &
        8 &
        71 &
        65
        \\
        
        \href{https://etherscan.io/block/21631012}{21631012} &
        134 &
        46 &
        88 &
        32 &
        95 (70.90\%) &
        6 (4.48\%) &
        37 (80.43\%) &
        59 (67.05\%) &
        110 &
        5 &
        35 &
        33
        \\
        
        \textbf{\href{https://etherscan.io/block/21631013}{21631013}} &
        175 &
        51 &
        124 &
        58 &
        118 (67.43\%) &
        7 (4.00\%) &
        44 (86.27\%) &
        74 (59.68\%) &
        136 &
        7 &
        68 &
        63
        \\
        
        \textbf{\href{https://etherscan.io/block/21631014}{21631014}} &
        200 &
        66 &
        134 &
        41 &
        137 (68.50\%) &
        6 (3.00\%) &
        50 (75.76\%) &
        87 (64.93\%) &
        157 &
        6 &
        70 &
        69
        \\
        
        \textbf{\href{https://etherscan.io/block/21631015}{21631015}} &
        138 &
        42 &
        96 &
        40 &
        104 (75.36\%) &
        5 (3.62\%) &
        40 (95.24\%) &
        64 (66.67\%) &
        118 &
        5 &
        25 &
        24
        \\
        
        \textbf{\href{https://etherscan.io/block/21631016}{21631016}} &
        180 &
        68 &
        112 &
        58 &
        107 (59.44\%) &
        24 (13.33\%) &
        37 (54.41\%) &
        70 (62.50\%) &
        121 &
        24 &
        378 &
        375
        \\
        
        \textbf{\textcolor{teal}{\href{https://etherscan.io/block/21631017}{21631017}}} &
        \textcolor{teal}{119} &
        \textcolor{teal}{38} &
        \textcolor{teal}{81} &
        \textcolor{teal}{43} &
        \textcolor{teal}{93 (78.15\%)} &
        \textcolor{teal}{5 (4.20\%)} &
        \textcolor{teal}{30 (78.95\%)} &
        \textcolor{teal}{63 (77.78\%)} &
        \textcolor{teal}{103} &
        \textcolor{teal}{5} &
        \textcolor{teal}{26} &
        \textcolor{teal}{26} 
        \\
        
        \textbf{\href{https://etherscan.io/block/21631018}{21631018}} &
        230 &
        100 &
        130 &
        46 &
        150 (65.22\%) &
        18 (7.83\%) &
        50 (50.00\%) &
        100 (76.92\%) &
        169 &
        18 &
        263 &
        255
        \\
        
        \textbf{\href{https://etherscan.io/block/21631019}{21631019}} &
        145 &
        47 &
        98 &
        30 &
        108 (74.48\%) &
        5 (3.45\%) &
        30 (63.83\%) &
        78 (79.59\%) &
        120 &
        5 &
        42 &
        40
        \\
        
        \textbf{\href{https://etherscan.io/block/21631020}{21631020}} &
        166 &
        55 &
        111 &
        39 &
        114 (68.67\%) &
        7 (4.22\%) &
        39 (70.91\%) &
        77 (69.37\%) &
        131 &
        6 &
        69 &
        62
        \\
        
        \hline\hline
        
        \textbf{Average} &
        \textbf{176} &
        \textbf{62} &
        \textbf{114} &
        \textbf{46} &
        \textbf{114} (\textbf{66.17}\%) &
        \textbf{12} (\textbf{6.92}\%) &
        \textbf{41} (\textbf{68.80}\%) &
        \textbf{74} (\textbf{66.06}\%) &
        \textbf{132} &
        \textbf{12} &
        \textbf{163} &
        \textbf{158} 
         \\
        \hline
    \end{tabular}%
    }
\vspace{-.5cm}
\end{table*}

\noindent\textbf{Observation-4.} \Cref{tab:ethereum_rb} highlights block-wise trends of the 20 recent blocks (from the timestamp: January 15, 2025, 04:13:23 PM UTC). As shown, most blocks have over 50\% independent transactions, with the highest parallelism in block 21631017, which has 119 transactions, of which 78.15\% are independent. The conflict chain is the shortest, with 5 transactions, accounting for 4.20\% of the block size. In particular, most conflicts are from contract transactions; out of 81 contract transactions, 18 (81-63) are conflicting, while only 8 (38-30) are conflicting from transfers. This block records the second-lowest number of conflicts (26) among the considered blocks. In contrast, block 21631010 has the least parallelism, with 48.31\% of transactions being conflicting, the conflict chain involving 16.43\% of the block size (207), and the highest number of conflicts (832). In summary, these blocks have an average of 176 transactions, of which 66.17\% are independent, and the conflict chain takes up 6.92\% of the block size. 

\noindent\textbf{Observation-5.} We also observed access skewness for these blocks (21631001--21631020), where a small number of EOAs and contract accounts are repeatedly accessed and emerge as hotspots for contention. The accounts  \href{https://etherscan.io/address/0x43506849d7c04f9138d1a2050bbf3a0c054402dd}{\texttt{0x4350...}} and \href{https://etherscan.io/address/0xc02aaa39b223fe8d0a0e5c4f27ead9083c756cc2}{\texttt{0xc02a...}} show strong access locality. Even relatively small blocks exhibit this skewness; for example, block 21631003, with 82 transactions, has over 5\% of accesses targeting a single EOA. Despite this, the conflict chains consist of a small fraction of the block size on average ($\sim$6.9\%). These observations suggest that \textit{access frequency and locality} are stronger indicators of the scope of parallel execution than raw transaction count.

\begin{figure}[!tb]
\centering
\pgfplotstableread[col sep=comma]{figs/results/RB_after.csv}\rbatable
\pgfplotstableread[col sep=comma]{figs/results/RB_before.csv}\rbbtable
\pgfplotstableread[col sep=comma]{figs/results/E2O_after.csv}\eoatable
\pgfplotstableread[col sep=comma]{figs/results/E2O_before.csv}\eobtable
\pgfplotstableread[col sep=comma]{figs/results/CK_after.csv}\ckatable
\pgfplotstableread[col sep=comma]{figs/results/CK_before.csv}\ckbtable

    \begin{tikzpicture}[scale = 1, font=\normalsize]
        \begin{axis}[
            base,
            title style = {at={(0.8,1.45)}, font = \small},
            width   = 2.15\linewidth,
            height  = 1.25\linewidth,
            enlarge x limits = 0.06,
            y label style	 = {at = {(-.125,.69)}},
            xlabel  = {Total Conflict Families},
            ylabel  = {Dense Conflict Family Size},
            legend style = {at = {(.2,1.43)}, anchor = north, legend columns = 1, fill = white},
        ]
            \addplot[
                black!80,
                only marks,
                mark = x,
                mark size = 5pt,
            ]
            table[
                x = {Conflict Families},
                y = {Most Danced Conflict Family Size}
            ]{\ckbtable};
            \addplot[
                black!80,
                only marks,
                mark = x,
                mark size = 5pt,
            ]
            table[
                x = {Conflict Families},
                y = {Most Danced Conflict Family Size}
            ]{\ckatable};
        
            \addplot[
                red,
                only marks,
                mark = +,
                mark size = 5pt,
            ]
            table[
                x = {Conflict Families},
                y = {Most Danced Conflict Family Size}
            ]{\eobtable};
            \addplot[
                red,
                only marks,
                mark = +,
                mark size = 5pt,
            ]
            table[
                x = {Conflict Families},
                y = {Most Danced Conflict Family Size}
            ]{\eoatable};
        
            \addplot[
                teal!90,
                only marks,
                mark = o,
                mark size = 3pt,
            ]
            table[
                x = {Conflict Families},
                y = {Most Danced Conflict Family Size}
            ]{\rbbtable};
            \addplot[
                teal!90,
                only marks,
                mark = o,
                mark size = 3pt,
            ]
            table[
                x = {Conflict Families},
                y = {Most Danced Conflict Family Size}
            ]{\rbatable};
        
            \legend{~E$_{ck}$~,,~E$_{e2}$~,,~E$_{rb}$~,}
        \end{axis}

        \pgfplotsset{
          hp inset/.style={
            axis lines = box,
            tick label style = {font = \tiny},
            clip = true,
          }
        }
    
        \begin{axis}[
          hp inset,
          xmax  = 550,
          ymax  = 550,
          xtick = {0, 100, 250, 350, 500},
          at    = {(rel axis cs:.514,.575)},
          width = 0.7\linewidth, height = 0.43\linewidth,
        ]
        \addplot[black!80, only marks, mark=x, mark size=2.2pt]
          table[x={Conflict Families}, y={Most Danced Conflict Family Size}] {\ckbtable};
        \addplot[black!80, only marks, mark=x, mark size=2.2pt]
          table[x={Conflict Families}, y={Most Danced Conflict Family Size}] {\ckatable};
        \end{axis}
        
        \begin{axis}[
          hp inset,
          xmax  = 550,
          ymax  = 550,
          xtick = {0, 100, 250, 350, 500},
          at    = {(rel axis cs:1.99,1.7)},
          width = 0.7\linewidth, height=0.43\linewidth,
        ]
        \addplot[red, only marks, mark=+, mark size=2.2pt]
          table[x={Conflict Families}, y={Most Danced Conflict Family Size}] {\eobtable};
        \addplot[red, only marks, mark=+, mark size=2.2pt]
          table[x={Conflict Families}, y={Most Danced Conflict Family Size}] {\eoatable};
        \end{axis}
        
        \begin{axis}[
          hp inset,
          xmax  = 550,
          ymax  = 550,
          xtick = {0, 100, 250, 350, 500},
          at    = {(rel axis cs:1.99,0.85)}, 
          width = 0.7\linewidth, height=0.43\linewidth,
        ]
        \addplot[teal!90, only marks, mark=o, mark size=2.2pt]
          table[x={Conflict Families}, y={Most Danced Conflict Family Size}] {\rbbtable};
        \addplot[teal!90, only marks, mark=o, mark size=2.2pt]
          table[x={Conflict Families}, y={Most Danced Conflict Family Size}] {\rbatable};
        \end{axis}

    \end{tikzpicture}
    \caption{Ethereum conflict family cluster structure: each point is a block; x is the number of conflict families and y is the size of the largest conflict family.}
    \label{fig:rw-family-cluster}
    \vspace{-.25cm}
\end{figure}
\begin{figure}[!tb]
\centering
\pgfplotstableread[col sep=comma]{figs/results/RB_after.csv}\rbatable
\pgfplotstableread[col sep=comma]{figs/results/RB_before.csv}\rbbtable
\pgfplotstableread[col sep=comma]{figs/results/E2O_after.csv}\eoatable
\pgfplotstableread[col sep=comma]{figs/results/E2O_before.csv}\eobtable
\pgfplotstableread[col sep=comma]{figs/results/CK_after.csv}\ckatable
\pgfplotstableread[col sep=comma]{figs/results/CK_before.csv}\ckbtable

    \begin{tikzpicture}[scale=1, font=\normalsize]
    
        \begin{axis}[
            base,
            xmin = 0,
            ytick  = {0, 10, 20, 30, 40, 50, 60},
            title style = {at={(0.8,1.45)}, font = \small},
            width=2.15\linewidth,
            height=1.25\linewidth,
            enlarge x limits = 0.06,
            y label style	 = {at = {(-.1,.69)}},
            xlabel={Longest Conflict Chain},
            ylabel={Block Size / Longest Conflict Chain},
            legend style = {at = {(.25,1.43)}, anchor = north, legend columns = 1, fill = white},
        ]
        
            \addplot[
                black!80,
                only marks,
                mark=x,
                mark size=4pt,
            ]
            table[
                x={Longest Conflict Chain},
                y={Block Size / Chain}
            ]{\ckbtable};
            
            \addplot[
                black!80,
                only marks,
                mark=x,
                mark size=4pt,
            ]
            table[
                x={Longest Conflict Chain},
                y={Block Size / Chain}
            ]{\ckatable};
            
            \addplot[
                red,
                only marks,
                mark=+,
                mark size=4pt,
            ]
            table[
                x={Longest Conflict Chain},
                y={Block Size / Chain}
            ]{\eobtable};
            
            \addplot[
                red,
                only marks,
                mark=+,
                mark size=4pt,
            ]
            table[
                x={Longest Conflict Chain},
                y={Block Size / Chain}
            ]{\eoatable};
            
            \addplot[
                teal!90,
                only marks,
                mark=o,
                mark size=4pt,
            ]
            table[
                x={Longest Conflict Chain},
                y={Block Size / Chain}
            ]{\rbbtable};
            
            \addplot[
                teal!90,
                only marks,
                mark=o,
                mark size=4pt,
            ]
            table[
                x={Longest Conflict Chain},
                y={Block Size / Chain}
            ]{\rbatable};
            
            \legend{~E$_{ck}$~,,~E$_{e2}$~,,~E$_{rb}$~,}
            
        \end{axis}

        \pgfplotsset{
          hp inset/.style={
            axis lines=box,
            tick label style={font=\tiny},
            clip=true,
          }
        }
    
        \begin{axis}[
          hp inset,
          xmax  = 500,
          ymax  = 55,
          xtick = {0, 100, 250, 350, 500},
          at    = {(rel axis cs:.514,.575)},
          width = 0.7\linewidth, height=0.43\linewidth,
        ]
        \addplot[black!80, only marks, mark=x, mark size=2.2pt]
          table[x={Longest Conflict Chain}, y={Block Size / Chain}] {\ckbtable};
        \addplot[black!80, only marks, mark=x, mark size=2.2pt]
          table[x={Longest Conflict Chain}, y={Block Size / Chain}] {\ckatable};
        \end{axis}
        
        \begin{axis}[
          hp inset,
          xmax  = 500,
          ymax  = 55,
          xtick = {0, 100, 250, 350, 500},
          at    = {(rel axis cs:1.99,1.7)},
          width = 0.7\linewidth, height=0.43\linewidth,
        ]
        \addplot[red, only marks, mark=+, mark size=2.2pt]
          table[x={Longest Conflict Chain}, y={Block Size / Chain}] {\eobtable};
        \addplot[red, only marks, mark=+, mark size=2.2pt]
          table[x={Longest Conflict Chain}, y={Block Size / Chain}] {\eoatable};
        \end{axis}
        
        \begin{axis}[
          hp inset,
          xmax  = 500,
          ymax  = 55,
          xtick = {0, 100, 250, 350, 500},
          at    = {(rel axis cs:1.99,0.85)}, 
          width = 0.7\linewidth, height=0.43\linewidth,
        ]
        \addplot[teal!90, only marks, mark=o, mark size=2.2pt]
          table[x={Longest Conflict Chain}, y={Block Size / Chain}] {\rbbtable};
        \addplot[teal!90, only marks, mark=o, mark size=2.2pt]
          table[x={Longest Conflict Chain}, y={Block Size / Chain}] {\rbatable};
        \end{axis}
    
    \end{tikzpicture}

    \caption{Ethereum theoretical parallelism bound as a function of longest conflict chain. Each point is a block; speedup is approximated by $\frac{\text{Block Size}}{\text{Longest Conflict Chain}}$. 
    }
    \label{fig:rw-speedup-vs-chain}
    \vspace{-.55cm}
\end{figure}
\noindent\textbf{Observation-6.} \Cref{fig:rw-family-cluster} and \Cref{fig:rw-speedup-vs-chain} shows clear differences between HPs in dependency structure and parallelism, while preserving a common underlying pattern. $E_{ck}$ periods exhibit a tighter concentration of blocks with fewer conflict families and shorter conflict chains, indicating relatively localized contention and higher parallelism. In contrast, $E_{2o}$ periods show a visibly broader spread, with a higher frequency of blocks containing large conflict families and long dependency chains, reflecting increased interaction with shared contracts and state hotspots. This shift results directly in the speedup bounds, where $E_{rb}$ periods exhibit a heavier tail with larger block sizes. In particular, although conflict families vary substantially between periods, the emergence of dominant families and the resulting conflict chains drive the effective serialization bottleneck in every HP. 
Overall, the comparison indicates a progressive tightening of parallelism, derived from growing state contention, while reinforcing that the conflict chain remains the principal metric for speedup. 



\noindent\textbf{Takeaway.} In Ethereum, over 50\% of the blocks in each HP contained more than 50\% independent transactions. The theoretical bound on speedup is constrained by the conflict chain, which accounts for $\sim$16\% of the block size on average across HPs. The change in independent transaction percentages over time and block-by-block, conflict chains, and conflict families indicates that no single parallel execution strategy is fit for all blocks. Moreover, HPs show significant shifts in conflict patterns, with contract transactions being the primary source of contention. Our observations highlight the need for an adaptive scheduling technique that dynamically chooses the best possible execution strategy and optimizes overall execution based on real-time block characteristics. Alternatively, a hybrid parallel execution model that leverages information available in transactions to derive conflicts with minimal overhead can maximize performance 
while minimize aborts and re-execution overhead.

\section{Solana's Empirical Study\label{sec:sol_analysis}}
Solana is the first \rwAware{} blockchain to support parallel execution. Transactions specify the states that can be read or written during execution. The Solana Sealevel~\cite{solanaSealevel} execute transactions in parallel using lock-based techniques (read-write locks) to identify independent transactions over multiple iterations~\cite{umbraresearch}. The conflict chain determines the iterations required for a block assuming a sufficient number of cores to fully exploit parallelism. 
We analyzed 1000 non-empty blocks from three distinct periods of Solana mainnet: the old historical period (S$_{ob}$) 61039000--61040210, the mid historical period (S$_{mb}$) 205465000--205466007, and the recent historical period (S$_{rb}$) 293971000--293972009. The Solana block consists of voting and non-voting transactions; we consider only non-voting transactions for our analysis.

\subsection{Analysis} As illustrated in \Cref{fig:solana_hp_analysis,fig:svm-family-cluster,fig:svm-speedup-vs-chain} and \Cref{tab:solana_rb}, the following observations can be made for real-world Solana blocks:

\begin{figure}[!b]
    \vspace{-.75cm}
    \scalebox{1.35}{\begin{tikzpicture}[scale = 1]
    \begin{axis}[
        ybar   = .7pt,
        xtick  = data,
        ymin   = 0,
        ymax   = 10000,
        ytick  = {0, 1000, 2000, 3000, 4000},
        width  = 15cm, 
        height = 10cm,
        enlarge x limits = 0.06,
        ylabel = {Transaction Count},
        y label style = {at = {(-.15,.39)}},
        xticklabel style = {rotate=35, anchor=east},
        bar width = 7pt,
        nodes near coords,
        every node near coord/.append style={font = \tiny, rotate = 90, xshift = 10pt, yshift = -2.8pt},
        legend style = {at = {(0.44,.81)}, anchor = north, legend columns = 3, draw = none, font=\tiny},
        symbolic x coords = {Block Size, Non-voting (NV) Txs, Successful NV Txs, Conflicting NV Txs, Independent NV Txs, Chain of Conflicting NV Txs, Conflict Families, Dense Conflict Family Size, Total Conflicts, Write-Write Conflicts},
        xticklabels = {Size, NVTs, SNVTs, CNVTs, INVTs, LCC, CF, DCFS, TCs, WWCs},
    ]
    
        \addplot[ybar, black!80, fill=black!80, every node near coord/.append style={yshift = -4.8pt}] 
        coordinates {
            (Block Size, 519) 
            (Non-voting (NV) Txs, 252) 
            (Successful NV Txs, 215) 
            (Conflicting NV Txs, 252) 
            (Independent NV Txs, 0) 
            (Chain of Conflicting NV Txs, 214) 
            (Conflict Families, 3) 
            (Dense Conflict Family Size, 233) 
            (Total Conflicts, 3664) 
            (Write-Write Conflicts, 3664)
        };
    
        \addplot[ybar, cyan!50, fill=cyan!50, every node near coord/.append style={yshift = -3pt}] 
        coordinates {
            (Block Size, 1972) 
            (Non-voting (NV) Txs, 113) 
            (Successful NV Txs, 78) 
            (Conflicting NV Txs, 101) 
            (Independent NV Txs, 12) 
            (Chain of Conflicting NV Txs, 52) 
            (Conflict Families, 19) 
            (Dense Conflict Family Size, 60) 
            (Total Conflicts, 886) 
            (Write-Write Conflicts, 676)
        };
    
        \addplot[ybar, purple, fill=purple, every node near coord/.append style={yshift = -.5pt}] 
        coordinates {
            (Block Size, 1249) 
            (Non-voting (NV) Txs, 334) 
            (Successful NV Txs, 182) 
            (Conflicting NV Txs, 310) 
            (Independent NV Txs, 23) 
            (Chain of Conflicting NV Txs, 119) 
            (Conflict Families, 39) 
            (Dense Conflict Family Size, 184) 
            (Total Conflicts, 3917) 
            (Write-Write Conflicts, 3751)
        };
    
        \legend{~S$_{ob}$~~, ~S$_{mb}$~~, ~S$_{rb}$}
    
        \node[anchor=north west] at (axis cs: Block Size, 10232) {
            \hspace{-.75cm}
            \resizebox{1\columnwidth}{!}{%
            \begin{tabular}{|r|c|c|c|}
                \hline 
                
                & 
                
                \textbf{\begin{tabular}[c]{@{}c@{}}~~\textcolor{black!80}{Old Historical}~~~\\\textcolor{black!80}{Period (S$_{ob}$)}\end{tabular}} 
                
                & 
                
                \textbf{\begin{tabular}[c]{@{}c@{}}~~\textcolor{cyan!80}{Mid Historical}~~~\\\textcolor{cyan!80}{Period (S$_{mb}$)}\end{tabular}} 
                
                & 
                
                \textbf{\begin{tabular}[c]{@{}c@{}}~~\textcolor{purple}{Recent Historical}~~~\\\textcolor{purple}{Period (S$_{rb}$)}\end{tabular}} 
                \\ \hline

                \textbf{Block Range}  & 
                61039000 - 61040210   & 
                205465000 - 205466007 & 
                293971000 - 293972009 \\\hline
                
                \textbf{Block Size} & 
                519  & 
                1972 & 
                1249 \\
                
                \textbf{Non-voting Txs (NVTs)} & 
                252 & 
                113 & 
                334 \\
                
                \textbf{Successful NV Txs (SNVTs)} & 
                215 & 
                78  & 
                182 \\
                
                \textbf{Conflicting NV Txs (CNVTs)} & 
                252 (100\%) & 
                101 (87\%)  & 
                310 (93\%) \\
                
                \textbf{Independent NV Txs (INVTs)} & 
                0 (0\%)   & 
                12 (13\%) & 
                23 (7\%) \\
                
                \textbf{Longest Conflict Chain NV Txs (LCC)} & 
                214 (84.28\%)& 
                52  (45.94\%)& 
                119 (36.50\%)\\
                
                \textbf{Conflict Families (CF)} & 
                3  & 
                19 & 
                39 \\
                
                \textbf{Dense Conflict Family Size (DCFS)} & 
                233 & 
                60 & 
                184 \\
                
                \textbf{Total Conflicts (TCs)} & 
                3664 & 
                886  & 
                3917 \\
                
                \textbf{Write-Write Conflicts (WWCs)} & 
                3664 & 
                676  & 
                3751 \\ \hline
            \end{tabular}%
            }
        };
    \end{axis}
\end{tikzpicture}}
    \caption{Solana's mainnet: analysis of 1000 blocks per historical period using read-write sets of non-voting transactions.
    }
    \label{fig:solana_hp_analysis}
\end{figure}
\begin{table*}
\centering
\caption{Block-wise trends of recent Solana blocks (314184230--314184249). The table reports transaction characteristics, independence ratios, conflict structure, and access contention derived from non-voting transaction. Despite large block sizes (with 477 non-voting transactions), only a small fraction of transactions remain independent ($\sim$4\% on average), while conflict chains and dense conflict families are consistently large across blocks. Several blocks exhibit very high conflicts (314184233 and~314184248), reflecting heavy contention on few accounts.
}

\label{tab:solana_rb}
\resizebox{\textwidth}{!}{%
    \begin{tabular}{|c|c|c|c|c|c|c|c|c|c|}
        \hline 
        \textbf{Block ID} & 
        \textbf{\begin{tabular}[c]{@{}c@{}} Block Size              \end{tabular}} & 
        \textbf{\begin{tabular}[c]{@{}c@{}} Non-voting (NV) Txs     \end{tabular}} & 
        \textbf{\begin{tabular}[c]{@{}c@{}} Successful NV-Txs       \end{tabular}} & 
        \textbf{\begin{tabular}[c]{@{}c@{}} Independent NV-Txs (\%) \end{tabular}} & 
        \textbf{\begin{tabular}[c]{@{}c@{}} Chain of Conflicts      \end{tabular}} & 
        \textbf{\begin{tabular}[c]{@{}c@{}} Conflict Families       \end{tabular}} & 
        \textbf{\begin{tabular}[c]{@{}c@{}} Dense Conflict Family Size \end{tabular}} & 
        \textbf{\begin{tabular}[c]{@{}c@{}} Total Conflicts         \end{tabular}} & 
        \textbf{\begin{tabular}[c]{@{}c@{}} W-W Conflicts           \end{tabular}} 
        \\ \hline \hline

        \textbf{\href{https://explorer.solana.com/block/314184230}{314184230}} & 2281 & 590 & 454 & 20 (3.39\%) & 386 & 32 & 527 & 16612 & 16145 \\
        
        \textbf{\href{https://explorer.solana.com/block/314184231}{314184231}} & 1537 & 518 & 170 & 22 (4.25\%) & 326 & 33 & 374 & 17116 & 17071 \\
        
        \textbf{\href{https://explorer.solana.com/block/314184232}{314184232}} & 2090 & 515 & 477 & 34 (6.6\%) & 326 & 46 & 407 & 8955 & 8871 \\
        
        \textbf{\textcolor{red}{\href{https://explorer.solana.com/block/314184233}{314184233}}} & \textcolor{red}{1689} & \textcolor{red}{335} & \textcolor{red}{298} & \textcolor{red}{2 (0.6\%)} & \textcolor{red}{238} & \textcolor{red}{6} & \textcolor{red}{301} & \textcolor{red}{5903} & \textcolor{red}{5566}  \\
        
        \textbf{\href{https://explorer.solana.com/block/314184234}{314184234}} & 1844 & 431 & 276 & 3 (0.7\%) & 226 & 6 & 419 & 8619 & 5762  \\
        
        \textbf{\href{https://explorer.solana.com/block/314184235}{314184235}} & 1304 & 72 & 40 & 6 (8.33\%) & 23 & 13 & 29 & 212 & 212  \\
        
        \textbf{\textcolor{teal}{\href{https://explorer.solana.com/block/314184236}{314184236}}} & \textcolor{teal}{2075} & \textcolor{teal}{446} & \textcolor{teal}{403} & \textcolor{teal}{41 (9.19\%)} & \textcolor{teal}{239} & \textcolor{teal}{59} & \textcolor{teal}{283} & \textcolor{teal}{4166} & \textcolor{teal}{3803}  \\
        
        \textbf{\href{https://explorer.solana.com/block/314184237}{314184237}} & 1722 & 330 & 304 & 11 (3.33\%) & 232 & 20 & 283 & 3184 & 3184  \\
        
        \textbf{\href{https://explorer.solana.com/block/314184238}{314184238}} & 1439 & 586 & 372 & 30 (5.12\%) & 375 & 42 & 519 & 4890 & 4739  \\
        
        \textbf{\href{https://explorer.solana.com/block/314184239}{314184239}} & 2026 & 517 & 377 & 17 (3.29\%) & 366 & 28 & 455 & 10900 & 5906  \\
        
        \textbf{\href{https://explorer.solana.com/block/314184240}{314184240}} & 2331 & 473 & 420 & 41 (8.67\%) & 266 & 54 & 355 & 6738 & 4073  \\
        
        \textbf{\href{https://explorer.solana.com/block/314184241}{314184241}} & 1863 & 498 & 262 & 16 (3.21\%) & 304 & 22 & 460 & 5614 & 4715  \\
        
        \textbf{\href{https://explorer.solana.com/block/314184241}{314184242}} & 1723 & 327 & 188 & 16 (4.89\%) & 197 & 23 & 295 & 2715 & 2102  \\
        
        \textbf{\href{https://explorer.solana.com/block/314184243}{314184243}} & 1579 & 430 & 192 & 6 (1.4\%) & 285 & 9 & 417 & 7672 & 5887  \\
        
        \textbf{\href{https://explorer.solana.com/block/314184244}{314184244}} & 2338 & 644 & 342 & 28 (4.35\%) & 299 & 38 & 533 & 12773 & 7727  \\
        
        \textbf{\href{https://explorer.solana.com/block/314184245}{314184245}} & 1568 & 544 & 348 & 19 (3.49\%) & 325 & 28 & 490 & 5018 & 4332  \\
        
        \textbf{\href{https://explorer.solana.com/block/314184246}{314184246}} & 2239 & 485 & 312 & 13 (2.68\%) & 314 & 21 & 436 & 11463 & 4036  \\
        
        \textbf{\href{https://explorer.solana.com/block/314184247}{314184247}} & 1484 & 477 & 233 & 16 (3.35\%) & 309 & 19 & 450 & 8480 & 4009  \\
        
        \textbf{\href{https://explorer.solana.com/block/314184248}{314184248}} & 2620 & 781 & 308 & 26 (3.33\%) & 320 & 38 & 394 & 15345 & 11245  \\
        
        \textbf{\href{https://explorer.solana.com/block/314184249}{314184249}} & 1711 & 535 & 210 & 11 (2.06\%) & 322 & 18 & 476 & 9763 & 7219 \\
        \hline\hline
        \textbf{Average} & \textbf{1873} & \textbf{477} & \textbf{299} & \textbf{19 (4\%)} & \textbf{284} & \textbf{28} & \textbf{395} & \textbf{8307} & \textbf{6330} \\
        \hline
    \end{tabular}%
}
\vspace{-.5cm}
\end{table*}

    \noindent\textbf{Observation-1.} As shown in~\Cref{fig:solana_hp_analysis}, the average block size has increased over 2$\times$ from the old-HP to the recent-HP; however, note that this increase is primarily due to voting transactions. Non-voting transactions have increased with a smaller margin, which has experienced a pronounced dip in the mid-HP relative to increased voting transactions. 

    \noindent\textbf{Observation-2.} The percentage of successful non-voting transactions has decreased over time, with the success rate of 85.32\% in the old-HP dropping to 69.03\% in the mid-HP and further to 54.50\% in the most recent historical period. This suggests that increasing network congestion may have contributed to transaction failures, potentially due to inaccuracies in transaction access specifications; the time at which specifications are generated by users and the time at which they are executed may differ due to intermediate ongoing execution at the validator nodes. The exact reasons for the increased transaction failures require further analysis, but failures may be due to network congestion or inaccuracies in the specifications. However, this indicates the limitations and efficiency of \rwAware{} execution models in high-contention workloads. 

    \noindent\textbf{Observation-3.} The percentage of independent transactions in Solana is considerably lower than that in Ethereum. However, there is a noticeable upward trend, with independent transactions increasing from 0\% in the old-HP (S$_{ob}$) to 7\% in the recent historical period (S$_{rb}$), while the mid-HP (S$_{mb}$) records 13\%. This suggests a gradual shift toward greater parallelism over time.

    Since Solana employs a lock-based multi-iteration parallel execution strategy, the number of conflict families has surged, from just 3 in S$_{ob}$ to 39 in S$_{rb}$, suggesting that despite high conflicts, multiple independent subsets of transactions can still be executed in parallel. Each subset is executed in parallel with the others, enhancing execution efficiency. The conflict chain of non-voting transactions has seen a substantial decline; it is reduced by 2.3$\times$. The conflict chain decreased from 84.92\% in S$_{ob}$ to 46.01\% in S$_{mb}$ and further to 35.62\% in S$_{rb}$ HP. The number of transactions within the most densely conflicted family has seen downward trends. This suggests more distributed conflicts and the possibility of improved parallel execution with more granular bottlenecks in transaction execution. This shows that transaction access patterns have changed over time, consequently improving the throughput of Solana's \rwAware{} execution model.

    \noindent\textbf{Observation-4.} Note that the majority of conflicts originate in write sets in historical blocks, accounting for 100\% in the old-HP, which decreases to 4.24\% (95.76\%) in recent blocks. This suggests that any approach that can minimize w-w conflicts could significantly enhance Solana's throughput. A potential solution is to adopt a multi-version data structure, similar to the one employed in \BSTM{}~\cite{blockstm}, which allows parallel execution while minimizing w-w contention.

    \noindent\textbf{Observation-5.}~\Cref{tab:solana_rb} highlights block-by-block analysis in recent blocks (chosen from the same period as Ethereum: January 15, 2025, 04:13:23 PM UTC). As shown, the size of the block varies by a significant margin, while the non-voting transactions is in the range of 72 to 781 per block, indicating increased voting activity in the network with more participating validators over past HPs. However, the independent transaction percentage varies from a minimum of $\sim$0.6\% in block 314184233 to a maximum of $\sim$9.19\% in block 314184236, with an average of 4\%, which is considerably lower compared to Ethereum. Additionally, in all these blocks, the majority of conflicts originate in write sets. The average conflict chain consists of 284 transactions, accounting for 59\% of the non-voting transactions, further emphasizing the possibility of w-w conflict optimization.

    \noindent\textbf{Observation-6.} We also observed strong access locality and skewness in the same blocks listed in \Cref{tab:solana_rb}, where a small set of accounts is repeatedly accessed. For example, the native {\href{https://explorer.solana.com/address/ComputeBudget111111111111111111111111111111}{\texttt{ComputeBudget1..}}} account appears in the majority of blocks (17 of 20) with high read accesses (722 in 314184248), while the \href{https://explorer.solana.com/address/CebN5WGQ4jvEPvsVU4EoHEpgzq1VV7AbicfhtW4xC9iM}{\texttt{Ce...iM}} and \href{https://explorer.solana.com/address/VN1RtW7sXqrSmRpfmgj1Hwyv2NnsimNmNsvZu5pcLnP}{\texttt{VN...nP}} accounts are written most frequently. Despite non-voting transactions varying significantly (from 72 in 314184235 to 781 in 314184248), access contention remains persistent. Moreover, blocks 314184240 and 314184249 exhibit large r-w sets with dependencies. Overall, these observations suggest that access patterns are skewed by the most frequently written accounts, which limit the effective parallelism of Solana's execution model, often more than block size or transaction count alone.
    \begin{figure}
        \centering
        \scalebox{1}{\pgfplotstableread[col sep=comma]{figs/results/HP3_RB.csv}\hprbtable
\pgfplotstableread[col sep=comma]{figs/results/HP2_Container.csv}\hpctable
\pgfplotstableread[col sep=comma]{figs/results/HP1_Old.csv}\hpotable

    \begin{tikzpicture}[scale = 1, font=\normalsize]
        \begin{axis}[
            base,
            title style = {at={(0.8,1.45)}, font = \small},
            ymax = 1000,
            xmax = 125,
            xtick  = {0, 25, 50, 75, 100, 125, 150},
            ytick  = {0, 250, 500, 750, 1000},
            width  = 2.15\linewidth,
            height = 1.25\linewidth,
            enlarge x limits = 0.06,
            y label style	 = {at = {(-.125,.69)}},
            xlabel = {Total Conflict Families},
            ylabel = {Largest Conflict Family Size},
            legend style = {at = {(1.35,1.4)}, anchor = north, legend columns = 1},
        ]
            \addplot[
                black!80,
                only marks,
                mark=star,
                mark size=3pt,
            ]
            table[
                x={Conflict Families},
                y={Most Danced Conflict Family Size}
            ]{\hpotable};
        
            \addplot[
                cyan!50,
                only marks,
                mark=square,
                mark size=2.5pt,
            ]
            table[
                x={Conflict Families},
                y={Most Danced Conflict Family Size}
            ]{\hpctable};
        
            \addplot[
                purple,
                only marks,
                mark=triangle,
                mark size=3pt,
            ]
            table[
                x={Conflict Families},
                y={Most Danced Conflict Family Size}
            ]{\hprbtable};
        
            \legend{~S$_{ob}$~,~S$_{mb}$~,~S$_{rb}$~}
        \end{axis}

        \pgfplotsset{
          hp inset/.style={
            axis lines=box,
            tick label style={font=\tiny},
            clip=true,
          }
        }
    
        \begin{axis}[
          hp inset,
          xmax  = 50,
          ymax  = 1000,
          xtick = {0, 10, 20, 30, 40, 50},
          at    = {(rel axis cs:.17,.575)},
          width = 0.6\linewidth, height=0.43\linewidth,
        ]
        \addplot[black!80, only marks, mark=star, mark size=2.2pt]
          table[x={Conflict Families}, y={Most Danced Conflict Family Size}] {\hpotable};
        \end{axis}
        
        \begin{axis}[
          hp inset,
          xmax  = 55,
          ymax  = 1000,
          ytick = \empty,
          xtick  = {0, 10, 20, 30, 40, 50},
          at    = {(rel axis cs:1.355,2.495)},
          width = 0.6\linewidth, height=0.43\linewidth,
        ]
        \addplot[cyan!50, only marks, mark=square, mark size=2.2pt]
          table[x={Conflict Families}, y={Most Danced Conflict Family Size}] {\hpctable};
        \end{axis}
        
        \begin{axis}[
          hp inset,
          xmax  = 125,
          ymax  = 1000,
          ytick = \empty,
          xtick = {0, 25, 50, 75, 100, 125},
          at    = {(rel axis cs:2.02,2.495)}, 
          width = 0.6\linewidth, height=0.43\linewidth,
        ]
        \addplot[purple, only marks, mark=triangle, mark size=2.2pt]
          table[x={Conflict Families}, y={Most Danced Conflict Family Size}] {\hprbtable};
        \end{axis}
    \end{tikzpicture}}
        \caption{Solana conflict family cluster structure: each point is a block; x is the number of conflict families and y is the size of the dense conflict family.}
        \label{fig:svm-family-cluster}
        \vspace{.2cm}
        \centering
        \scalebox{1}{\pgfplotstableread[col sep=comma]{figs/results/HP3_RB.csv}\hprbtable
\pgfplotstableread[col sep=comma]{figs/results/HP2_Container.csv}\hpctable
\pgfplotstableread[col sep=comma]{figs/results/HP1_Old.csv}\hpotable

    \begin{tikzpicture}[scale=1, font=\normalsize]
        \begin{axis}[
            base,
            xmin = 0,
            xtick  = {0, 200, 400, 600, 800, 1000},
            ytick  = {0, 5, 10, 15, 20},
            title style = {at={(0.8,1.45)}, font = \small},
            width=2.15\linewidth,
            height=1.25\linewidth,
            enlarge x limits = 0.06,
            y label style	 = {at = {(-.1,.69)}},
            xlabel={Longest Conflict Chain},
            ylabel={Non-Voting Txs / Longest Conflict Chain},
            legend style = {at = {(1.35,1.41)}, anchor = north, legend columns = 1},
        ]
        
            \addplot[
                black!80,
                only marks,
                mark=star,
                mark size=3pt,
            ]
            table[
                x={Longest Conflict Chain},
                y={Parallelism Factor (PF): Block Size / Chain}
            ]{\hpotable};
            
            \addplot[
                cyan!50,
                only marks,
                mark=square,
                mark size=3pt,
            ]
            table[
                x={Longest Conflict Chain},
                y={Parallelism Factor (PF): Block Size / Chain}
            ]{\hpctable};
            
            \addplot[
                purple,
                only marks,
                mark=triangle,
                mark size=3pt,
            ]
            table[
                x={Longest Conflict Chain},
                y={Parallelism Factor (PF): Block Size / Chain}
            ]{\hprbtable};
            
            \legend{~S$_{ob}$~,~S$_{mb}$~,~S$_{rb}$~,}
            
        \end{axis}

        \pgfplotsset{
          hp inset/.style={
            axis lines=box,
            tick label style={font=\tiny},
            clip=true,
          }
        }
    
        \begin{axis}[
          hp inset,
          xmax  = 1000,
          ymax  = 20,
          ytick  = {0, 5, 10, 15, 20},
          xtick = {0, 200, 400, 600, 800},
          at    = {(rel axis cs:.17,.5675)},
          width = 0.6\linewidth, height=0.43\linewidth,
        ]
        \addplot[black!80, only marks, mark=star, mark size=2pt]
          table[x={Longest Conflict Chain}, y={Parallelism Factor (PF): Block Size / Chain}] {\hpotable};
        \end{axis}
        
        \begin{axis}[
          hp inset,
          xmax  = 1000,
          ymax  = 20,
          ytick = \empty,
          xtick = {0, 200, 400, 600, 800},
          at    = {(rel axis cs:1.355,2.495)},
          width = 0.6\linewidth, height=0.43\linewidth,
        ]
        \addplot[cyan!50, only marks, mark=square, mark size=2pt]
          table[x={Longest Conflict Chain}, y={Parallelism Factor (PF): Block Size / Chain}] {\hpctable};
        \end{axis}
        
        \begin{axis}[
          hp inset,
          xmax  = 1000,
          ymax  = 20,
          ytick = \empty,
          xtick = {0, 200, 400, 600, 800},
          at    = {(rel axis cs:2.02,2.495)},
          width = 0.6\linewidth, height=0.43\linewidth,
        ]
        \addplot[purple, only marks, mark=triangle, mark size=2pt]
          table[x={Longest Conflict Chain}, y={Parallelism Factor (PF): Block Size / Chain}] {\hprbtable};
        \end{axis}
    
    \end{tikzpicture}}
        \caption{Solana theoretical parallelism bound as a function of longest conflict chain. Each point is a block; speedup is approximated by $\frac{\text{\# Non-Voting Transactions}}{\text{Longest Conflict Chain}}$. 
        }
        \label{fig:svm-speedup-vs-chain}
        \vspace{-.35cm}
    \end{figure}

    \noindent\textbf{Observation-7.} \Cref{fig:svm-family-cluster,fig:svm-speedup-vs-chain} characterizes the dependency structure and parallelism limits of recent Solana blocks. In contrast to Ethereum, Solana exhibit a small number of conflict families that quickly grow into dense families, indicating contention concentrated on a few dominant program-level or state accounts rather than being distributed across many independent groups. This behavior in $S_{rb}$ is more explicit, where dense conflict families persist even when the total number of families is modest. As shown in \Cref{fig:svm-speedup-vs-chain}, this dependency structure directly limits the maximum achievable speedup: the conflict chain scales rapidly with the number of non-voting transactions, leading to a sharp reduction in the theoretical speedup bound. Compared to Ethereum, where conflicts are more fragmented and shorter chains often preserve unexploited parallelism, Solana shows a tighter trend between workload size and serialization depth, resulting in consistently lower parallel speedup bounds despite higher throughput. 

\noindent\textbf{Takeaway.} With these observations, we conclude that independent transactions remain very low on the Solana network in all three HPs, with an average of $\sim$4\% in recent blocks, while w-w conflicts dominate and account for $\sim$76.2\% of all conflicts within a given block. The conflict chain has decreased significantly, from 84.92\% to 35.62\%, from the older HP to the recent HP, but conflict families have increased from 3 to 39, indicating more granular conflicts and increased parallelism. Furthermore, the success rate of non-voting transactions has dropped from $\sim$85.32\% to $\sim$54.50\%, highlighting the need for adaptive or hybrid-execution strategies that efficiently exploit access specifications to improve throughput.

\section{Conflicts in Ethereum versus Solana\label{sec:eth_vs_solana}}

As shown in \Cref{fig:eth_vs_solana_analysis}, Solana's larger block size (352, non-voting transactions) for recent 1000 blocks compared to Ethereum's 177 highlights a fundamentally different architecture. Solana supports high throughput and parallel execution, leading to higher raw data per block due to its \rwAware{} model. Ethereum, in contrast, structures its blocks around finalized, gas-accounted transactions, reflecting a design optimized for decentralization and efficiency. 

This distinction extends to conflict behavior: Ethereum exhibits a higher percentage of independent transactions (112, 63.27\%) compared to Solana (29, 8.24\%), reflecting a more modular structure. 
Ethereum also has a higher number of conflict families (130) compared to Solana (48), showing more isolated contention hotspots. In contrast, Solana's design, optimized for high throughput, results in longer conflict chains (109, 30.97\%) compared to Ethereum (15, 8.47\%), showing deeper transaction interdependencies. Additionally, Solana has fewer conflict families but dense conflict families (176, 50\%), which are much denser than those in Ethereum (15, 8.47\%), suggesting high contention by specific applications.

We also analyze the parallelism factor ($\text{\footnotesize PF=}\frac{\text{block size}}{\text{longest conflict chain}}$), which represents the theoretical upper bound on parallel execution derived from block size and the conflict chain, where higher values indicate more parallelism. The effective parallel width ($\text{\footnotesize EPW =}\frac{\text{independent transactions}}{\text{longest conflict chain}}$) captures how much of this potential can actually be exploited in practice, with higher values indicating better utilization of available parallelism, while the conflict density ($\text{\footnotesize CD=}\frac{\text{conflicting transactions}}{\text{block size}}$) measures the fraction of transactions involved in conflicts, and conflict family fragmentation ($\text{\footnotesize CFF=}\frac{\text{conflict families}}{\text{conflicting transactions}}$) captures how evenly conflicts are distributed across independent families, with higher values indicating better fragmentation and greater parallelism. The scheduling sensitivity score ($\text{\footnotesize SSS=}\frac{\text{dense conflict family size}}{\text{longest conflict chain}}$) reflects the degree to which execution is dominated by a single dense conflict family, where higher values indicate tighter serialization and greater sensitivity to scheduling decisions. 

\begin{figure}
    \centering
    \scalebox{1.05}{\begin{tikzpicture}[scale = 1.15, font = \small]
    \begin{axis}[
        base,
        ybar   = 2pt,
        xtick  = data,
        ymin   = 0,
        ymax   = 450,
        ytick  = {0, 100, 200, 300, 400},
        width  = 15cm, 
        height = 8cm,
        enlarge x limits = 0.06,
        ylabel           = {Transaction Count or Value},
        y label style	 = {at = {(-0.15,.7)}},
        xticklabel style = {rotate=45, anchor=east},
        title={\shortstack[l]{
        \textbf{IndTxs}: independent transactions\\
        \textbf{LCC}: longest conflict chain\\
        \textbf{CF}: conflict families\\
        \textbf{DCFS}: dense conflict family size\\
        \textbf{PF}: parallelism factor\\
        \textbf{EPW}: effective parallelism width\\
        \textbf{CD}: conflict density\\
        \textbf{CFF}: conflict family fragmentation\\
        \textbf{SSS}: scheduling sensitivity score
        }},
        title style={at={(1.15,.51)}, font = \scriptsize},
        bar width = 12pt,
        nodes near coords,
        every node near coord/.append style={rotate = 90, font = \scriptsize},
        legend style = {at = {(0.45,1.25)}, anchor = north, legend columns = 1, fill =none},
        symbolic x coords = {Block Size, IndTxs, LCC, CF, DCFS, PF, EPW, CD, CFF, SSS},
    ]
    
    \addplot+[teal!60, fill = teal!60, every node near coord/.append style={yshift = -8pt, xshift = 14pt}] 
    coordinates {
        (Block Size, 177) 
        (IndTxs, 112) 
        (LCC, 15) 
        (CF, 130) 
        (DCFS, 15)
        (PF, 16.72)
        (EPW, 11.17)
        (CD,0.36)
        (CFF,2.30)
        (SSS,0.99)
    };
    
    \addplot+[orange!65!yellow!90, fill = orange!65!yellow!90, every node near coord/.append style={yshift = -3.5pt, xshift = 18pt}] 
    coordinates {
        (Block Size, 352) 
        (IndTxs, 29) 
        (LCC, 109) 
        (CF, 48) 
        (DCFS, 176)
        (PF, 3.28)
        (EPW, 0.359)
        (CD,.073)
        (CFF,0.587)
        (SSS,1.55)
    };
    
    \legend{~Ethereum~, ~Solana~}
    \end{axis}
\end{tikzpicture}}
    \caption{Ethereum versus Solana in 1000 recent historical blocks.}
    \vspace{-.2cm}
    \label{fig:eth_vs_solana_analysis}
    \vspace{-.25cm}
\end{figure}

Ethereum exhibits higher PF (16.72) and EPW (11.17) than Solana's 3.28 and 0.36, respectively, indicating substantial parallel execution opportunities. Although Solana shows a lower CD (0.07 compared to Ethereum's 0.36), its conflicts are highly concentrated rather than fragmented. This is reflected in a much lower CFF 0.59 compared to Ethereum's 2.3, indicating that conflicts in Solana cluster into a few dense families. As a result, Solana exhibits a higher SSS (1.55) than Ethereum (0.99), indicating that execution is largely constrained by a single dominant conflict family. In contrast, Ethereum's higher CFF and lower SSS indicate reduced scheduling sensitivity and higher parallelism.

In summary, it is crucial to understand conflict metrics for efficient parallel execution. Both chain exhibit significant parallel execution potential in different execution models; they differ in key aspects of conflicts, available parallelism, and potential hotspots. Ethereum, with lower conflict rates, could result in higher parallelism, showing potential for execution efficiency and lower abort rates. Despite having lower execution throughput than Solana today, Ethereum could achieve substantially higher throughput under parallel block execution. Solana exhibits higher conflict rates, particularly due to the large number of w-w conflicts, resulting in more granular congestion, highlighting the limitations of its current execution model. While high parallelism factor alone is not sufficient, Ethereum shows potential parallelism through fragmented conflicts, whereas Solana concentrated contention structure tightly bounds effective parallel width. 
These patterns, block composition, transaction independence, and conflict complexity, illustrate how Ethereum and Solana make different trade-offs in their pursuit of efficient execution. However, both chains could be further optimized 
to improve execution efficiency.


Despite Solana's higher transaction throughput on its mainnet compared to Ethereum, it faces the challenge of increasingly common transaction failures. Likewise, since Ethereum still executes transactions sequentially, there is ongoing research in parallel EVM~\cite{pevm,monad,seiparallelexe,polygonParallelization,iBTM_AFT_2025} inspired by STMs that could handle contention more effectively. Given current trends, we believe that both chains (including other EVM- and SVM-based chains) would benefit from adaptive and hybrid scheduling techniques to exploit parallelism with evolving workload characteristics and user behavior. Solana, in particular, requires innovations to mitigate w-w conflicts, potentially through the adoption of multi-version data structures, while Ethereum needs to implement parallel execution for the EVM.

\section{Concluding Remarks\label{sec:conc}}
Blockchains typically process transactions in a strict order, one after another, to ensure that all nodes reach identical states. This approach guaranties consistency, but limits throughput. Identifying which transactions truly depend on each other, i.e., which transactions conflict, can help maximize throughput. The EVM processes transactions sequentially, without advance knowledge of what state each transaction will access. The SVM, on the other hand, requires clients to specify upfront what states a transaction will read or write. 

Our analysis of Ethereum blocks revealed something interesting: consistently across all time periods, more than 60\% of the transactions are completely independent and could theoretically be executed in parallel.
In recent 10k blocks, 63.27\% of transactions are independent. The \emph{conflict chain}, i.e., transactions that must run sequentially, is only about 8-9\% of the block size. 
The conflict rate for Ethereum contract transactions ($\frac{115-73}{115} \approx 0.36$) is the same as for ETH transfers ($\frac{63-40}{63} \approx 0.36$). Additionally, more than 99\% of recent blocks have at least 40\% independent transactions. In contrast, Solana blocks show different conflict patterns. From the old to the recent historical period, the conflict chain shortened (84.92\% to 35.62\%) and conflict families increased (3 to 39), indicating a shift toward fine-grained parallelism. However, recent blocks show consistently low independence (avg. $\sim$4\%), with write-write conflicts comprising most contention ($\sim$76\%). Additionally, the success rate of non-voting transactions dropped from $\frac{215}{252} \approx 0.85$ to $\frac{299}{477} \approx 0.62$, showing the limitations of Solana's current \rwAware{} scheduling under high contention workloads.

This indicates that block-to-block variation is substantial, necessitating adaptive execution strategies. Moreover, our analysis confirms our initial hypothesis: traditional sequential execution leaves a large amount of performance potential untapped, which can be effectively exploited through accurate conflict detection and parallel execution in both EVM and SVM. The \rwAware{} and \rwOblivious{} models each exhibit distinct strengths and limitations and neither dominates all workloads or conflict patterns. While parallel execution is essential for both Ethereum and Solana, their fundamentally different conflict models imply that maximizing parallelism requires distinct, often hybrid or adaptive execution strategies, rather than a single universal approach.


\bibliographystyle{IEEEtran}
\bibliography{references}

\end{document}